\documentclass[10pt, preprint]{aastex}

\usepackage{graphics}
\usepackage[hyperfootnotes=false]{hyperref}
\usepackage{color} 
\usepackage{amsmath}

\newcommand{\bea}{\begin{eqnarray}}
\newcommand{\eea}{\end{eqnarray}}
\def\beq{\begin{equation}}
\def\eeq{\end{equation}}

\def\r{{\it RHESSI}\ }
\def\f{{\it Fermi}\ }

\def\a{{\alpha}}
\def\b{{\beta}}

\def\d{{\delta}}
\def\g{{\gamma}}
\def\e{\epsilon}

\def\he3{$^3$He\,}
\def\he4{$^4$He\,}
\def\coul{{\rm Coul}}

\def\tsc{\tau_{\rm sc}\,}
\def\tac{\tau_{\rm ac}\,}
\def\tdiff{\tau_{\rm diff}\,}
\def\tcross{\tau_{\rm cross}\,}
\def\tesc{T_{\rm esc}\,}

\def\tloss{\tau_{\rm L}\,}

\def\tcoul{\tau_\coul}

\begin{document}

\title{Particle Acceleration in Solar Flares and Associated CME Shocks}
\author{Vah\'e Petrosian$^{1,2}$}
\affil{$^{1}$Department of Physics and KIPAC, Stanford University,
Stanford, CA 94305, USA\\ 
$^{2}$Department of Applied Physics, Stanford University, Stanford, CA 94305,
USA\\ 
}
\shorttitle{Particle Acceleration in Solar Flares}
\shortauthors{Petrosian}
%\maketitle

\begin{abstract}

Observations relating the  characteristics of
electrons seen near Earth (SEPs) and those producing flare radiation show that
in certain (prompt) events the origin of both population appears to be the flare
site, which show strong correlation between the number and spectral
index of SEP and hard X-ray radiating electrons, but in others(delayed),
which are associated with fast CMEs, this relation is  complex  and
SEPs tend to be harder. Prompt event spectral relation disagrees with that
expected in  thick or thin target models. We show that using a  a more accurate
treatment of the 
transport of the accelerated electrons to the footpoints and to the Earth can
account for this discrepancy. Our results are consistent with those found by
Chen \& Petrosian (2013) for two flares using non-parametric inversion methods,
according to which  we have weak diffusion conditions, and trapping 
mediated by magnetic field convergence. The weaker correlations and harder
spectra of delayed events can come about by re-acceleration of electrons in the
CME shock environment. We describe under what conditions such a hardening can be
achieved. Using this (acceleration at the flare and re-acceleration in the CME)
scenario we show that we can describe the similar dichotomy that exists between
the so called impulsive, highly enriched ($^3$He and heavy ions) and softer SEP
events, and stronger more gradual SEP events with near normal ionic abundances
and harder spectra. These methods can be used to distinguish  the acceleration
mechanisms and to constrain their characteristics.

\end{abstract}


\section{INTRODUCTION}
\label{intro}

Solar flares accelerate electrons and ions with varied characteristics deduced
either from direct observations of the intensity, spectrum and composition of
solar
energetic particles (SEPs) by near Earth instruments or indirectly from the
radiation they produce while interacting with background
solar particles and fields. Electrons produce hard X-rays (HXR) via nonthermal
bremsstrahlung (NTB) and microwaves via synchrotron mechanisms in the lower
solar atmosphere and type III (and other radio) bursts in the upper corona and
beyond.
Accelerated protons (and other ions) interacting with background ions result in
de-excitation nuclear lines in the 1 to 7 MeV range, neutrons, and pions which
decay
into (mainly) $>70$ MeV gamma-rays. However, most of the energy of $<\sim 100$
MeV
electrons and $<\sim 1$ GeV protons goes into heating and evaporation of the
flare
plasma which then gives rise to thermal radiation from soft X-rays (SXR) to
sub-millimeter range. There are  two possible acceleration sites; one
in the low solar corona, where reconnection seems to be the energizing
mechanism, and the other  at
higher corona and up to several solar radii from the Sun in the environment of
the coronal mass ejection (CME), possibly by a shock  and
mediated by turbulence. An important question
is whether radiating particles and SEPs are accelerated at only one of these
or at both sites, and if there are
observations which can answer this question and shed light on the
characteristics of the two acceleration mechanisms. This is the main goal
of the work presented here.

In the case of SEP ions this dichotomy has manifested itself in the differences
between
shorter duration or more impulsive events which show large enhancement of
$^3$He and heavy ions and longer duration or more gradual events with
photospheric abundances and harder spectra (see e.g.  Mason et al. 2000 \&
2002; Reames et al. 1994, 1997; Ng \& Reames 1994. For a more recent review see
Reames 2013). The former
appear to originate from the flare site and stochastic acceleration by
turbulence can account for their softer spectra and extreme enhancement of
$^3$He  (Liu et al. 2004, 2006). The gradual events, in addition to being
harder, tend to be stronger and have near normal photospheric abundances.
Stochastic acceleration by turbulence has often been considered to be the
mechanism of production of these SEPs (for observational evidence see e.g. Mason
et al. 1986;
Mazur et al. 1992; and theoretical interpretations see Miller et al. 1996;
Miller 2003). However, since the gradual events are
often associated with fast CMEs, and it is generally believed  that the CME
driven shock plays a major role in acceleration of these SEPs (see e.g. Zank
2012 Jokipii \& Giacalone 1996). As described
in \S 3, SEP electron
observations also point out to presence of more impulsive type flares with close
association with type III bursts and other flare emissions, and gradual flares
with a less defined relation with radiative signatures and  strong association
with proton rich SEPs (Krucker et al. 1999; Haggetry \& Roelof 2002; Maia
\& Pick 2004;  Klein et al. 2005; Krucker et al. 2007).

High spatial resolution observations  in the HXR regime by {\it Yohkoh};
(Masuda et al. 1994;  Petrosian et al. 2002) and by the {\it Ramaty High
Energy Solar Spectroscopic Imager} (\r) (Lin et al. 2002; Krucker \& Lin
2008; Liu et al. 2008),
and few cases in gamma-rays by \r (Hurford et al. 2003) and by the {\it Fermi
Gamma-ray
Observatory} (\f) (Ajello et al.  2013) have established that the nonthermal
radiations produced by both electrons and ions originate in  and below the
corona. Typically HXR and gamma-ray producing particles
must traverse
column depths of $> 10^{20}$ and $> 10^{24}$ cm$^{-2}$, respectively,
which are reached below the transition region and below the photosphere. It
is
generally believed that electrons that produce the microwaves and HXRs are
accelerated in or near the reconnection region in the low corona, possibly just
above the top of the flaring loops. These emissions 
normally are limited to the impulsive phase of a flare that lasts few
seconds to several minutes. The impulsive HXR emissions are sometimes
associated with high energy SEP electrons observed near the Earth (see e.g.
Wang et al. 2012).

 The relation of the characteristics of these electrons with those producing the
HXRs  depends on whether we are dealing with impulsive-prompt or
gradual-delayed SEP events. The differences between these two populations 
can also shed light on the acceleration site and mechanism. This will  be the
primary focus of this paper. 

The seed particles accelerated at the reconnection site in the low corona are
expected to be the background plasma particles but the
charcteristics of
seed particles to be accelerated by the shock in the gradual events are not 
known. In this paper we  explore a scenario where the flare
accelerated ions and electrons are the
seeds that are re-accelerated in the CME environment of the
gradual-delayed events. 

In general, gamma-ray observations, in particular
those by {\it Fermi}, point to a more complicated picture than that derived
from HXR
observations. Flare emitted
gamma-rays were observed first by the gamma-ray spectrometer (GRS) on board the
{\it
Solar Maximum Mission} (SMM) (Chupp et al. 1982) and later by the instruments,
notably EGRET, on the {\it Compton Gamma-ray Observatory} (CGRO), showing  both
impulsive and long duration de-excitation lines and pion decay emissions (see
review by Chupp \& Ryan 2009). But observations by the \f gamma-ray burst
monitor (GBM) and the large area telescope (LAT), with a superior sensitivity
and spatial resolution, have provided new insights and raised new questions.
Some
flares, like the June 12, 2011 (see Ackermann et al. 2012) show  simultaneous
HXR and gamma-ray  impulsive emissions indicating acceleration of both electrons
and ions in the low corona. But, as shown in Ackermann et al. (2014), a
majority of \f flares show long duration ($>100$ MeV)emission lasting sometimes
over 10
hours after the
impulsive phase. There does not seem to be any other associated  emissions
during
the extended phase, but most of these flares are associated with fast CMEs and
significant flux of SEPs. Irrespective of whether the extended emission is
NTB by relativistic electrons or from decay of pions produced by
$>320$ MeV protons (and other ions), acceleration in the CME shock may  be at
work here as well. This possibility has gained further support from \f
observations of
significant impulsive and extended gamma-ray emission from several flares
located behind the limb (BTL) as deduced from {\it STEREO} observations
(Pesce-Rollins et al. 2015). 

In the next section we describe  the details of the combined model of
acceleration in the corona and re-acceleration at the shock. We  apply this
to explain the relation between impulsive and gradual events as observed
for electrons in \S 3 and He ions in \S 4. A brief summary and discussion is
given in \S 5.

\section{COMBINED FLARE AND CME SHOCK ACCELERATION MODEL}
\label{sec:model}

Our basic model for acceleration is shown in Figure 1. The  left panel shows a
cartoon of the model for acceleration, transport and radiation at  the
X-reconnection site in the corona.  Acceleration can be by the 
second order Fermi (or stochastic) acceleration by turbulence (see e.g.
Petrosian \& Liu 2004), by a standing shock produced by the down flow from
the  X-reconnection site (see e.g. Guo \& Giacalone 20120), or in the merging
of
islands shown to arise in PIC simulations  during reconnection (Drake et al.
2006 \& 2013; Le et al. 2012; Oka et al. 2013). The latter model
has been invoked as a possible mechanism in the downstream of the CME shock (le
Roux et al. 2015; Zank et al; 2015). The right panel shows a
similar cartoon for acceleration in the
environment of the CME. The (red) rectangles shows the cross
section of the box within which particles are accelerated. For our purposes here
the relevant parameters of the acceleration and transport of particles are the
momentum ($p$) and pitch angle cosine ($\mu$) diffusion rates $D_{pp}/p^2$ and
$D_{\mu\mu}$, and direct
energy gain $A(E)$ and loss ${\dot E}_L$ rates, where $E$ is the particle
energy.%
\footnote{Here and in what follows we neglect the effects of the third
diffusion coefficient $D_{\mu p}$ which are generally small (see
Schlickeiser 1989; Petrosian \& Liu 2004)}.
For an isotropic
pitch angle distribution or for the gyro phase and pitch angle
averaged particle energy distribution $f(t,s,E)$, where $s$ is distance along
field lines, integrated over the volume $V$ of the
acceleration site with cross section area ${\cal A}(s)$, namely $N(t,E)=\int
{\cal
A}(s)dsf(t,s,E)$, is governed by the
so-called leaky-box model kinetic equation (see Petrosian 2012)

\beq   %\vspace{-0.2cm}
\label{KE} 
{\partial N \over \partial t} 
 = {\partial \over \partial E} \left(D_{EE}{\partial N \over \partial E}\right) 
 - {\partial \over \partial E} [(A - \dot E_{\rm L}) N] 
 - {N \over T_{\rm esc}} +{\dot Q},  
\eeq 
where ${\dot Q}$ is a source term, $D_{EE}=v^2D_{pp}$,  and  the direct
acceleration
coefficients  for stochastic and shock accelerations are given as
\beq 
A_{\rm SA}=\xi {D_{EE}\over E},\,\,\,\, {\rm and}\,\,\,\, 
A_{\rm sh}=\zeta E\left({u^2_{\rm sh}\over 4\kappa_{ss}}\right)=\zeta
E\left({u_{\rm
sh}\over
v}\right)^2\left(\langle{(1-\mu^2)^2\over {\bar
D}_{\mu\mu}}\rangle \right)^{-1}.
\label{accrates}
\eeq 
Here $u_{\rm sh}$ is the shock velocity, $\kappa_{ss}$ is the coefficient of
spatial diffusion (along the field
lines), $\xi=(2\gamma^2 -1)/(\gamma^2+\gamma)$, $\g$ is the Lorentz factor, and
$\zeta$ depends on the shock compression ratio and other factors 
(see, e.g.~Steinacker et al.~1988). The energy loss rate at low energies (for
both electrons and ions) is dominated by Coulomb collisions  with background 
particles (mainly electrons).%
\footnote{Coulomb collision also cause  pitch angle scattering, and therefore
spatial diffusion along the field lines, with a rate that is comparable to
energy
loss rate at nonrelativistic energies but decreases as $1/(\g^2-1)$.
They also cause energy diffusion which is negligible in the cold target
case ($E\gg kT$), but can be comparable to the energy loss rate for a hot target
as  $E$ approaches $kT$ (see Petrosian \& Kang 2015), where $T$ is the
background plasma temperature.}
At higher energies (not relevant for the discussion here) inelastic
interactions  (synchrotron, inverse 
Compton and bremsstrahlung) for electrons and  (nuclear line excitation, neutron
and pion productions) for ions become important. 

\begin{figure}[!ht]
\begin{center}
\includegraphics[width=8.0cm]{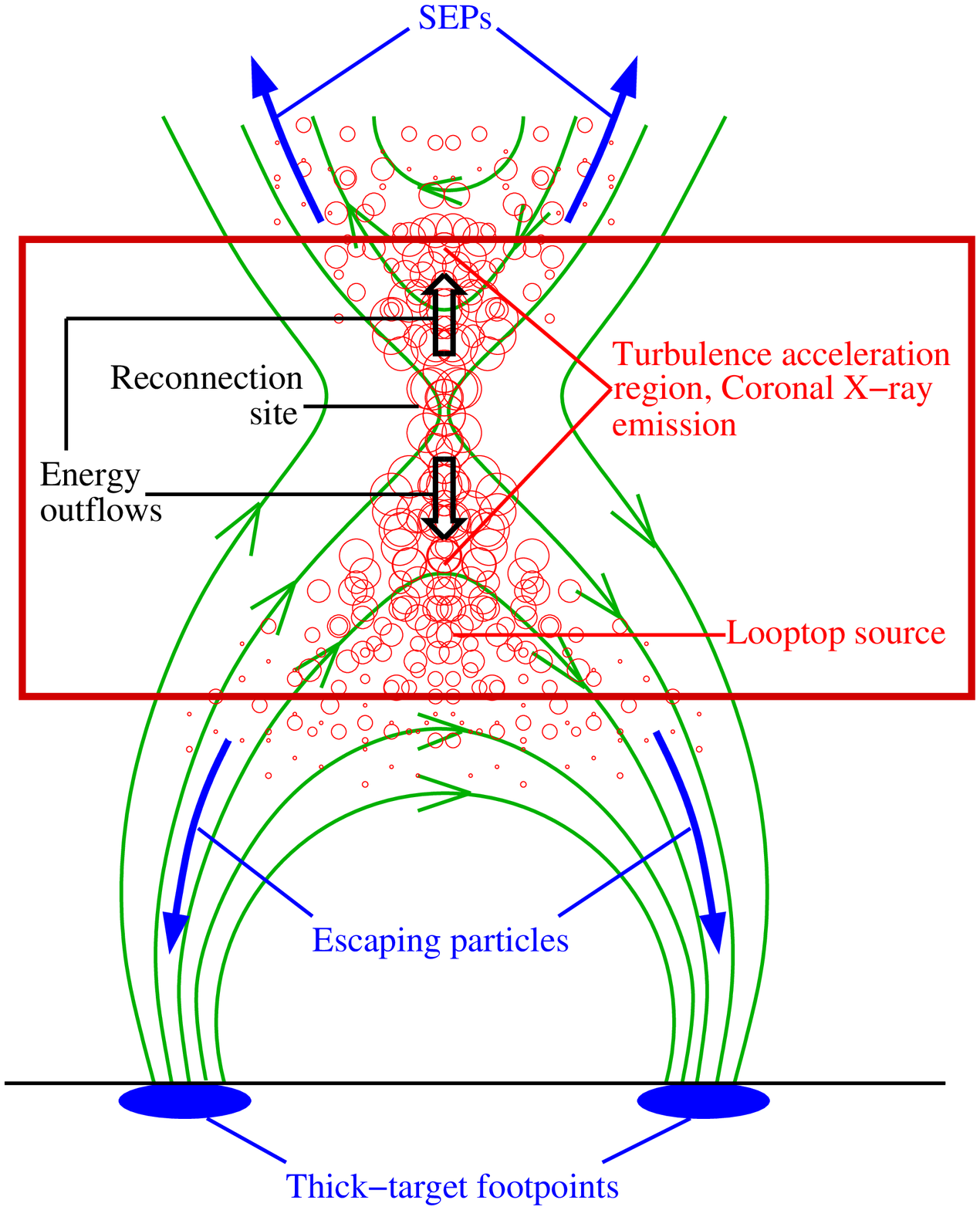}
\includegraphics[width=8.0cm]{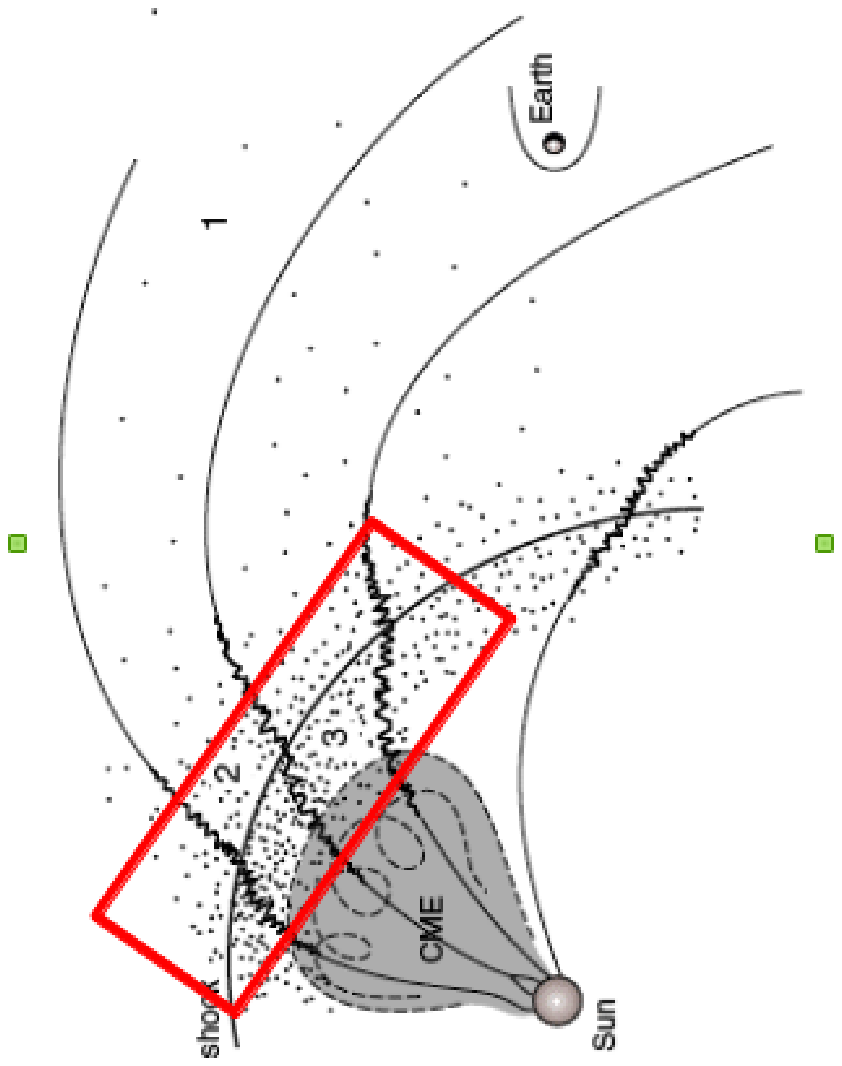}
\caption{{\bf Left:} A schematic representation of the reconnecting field
(solid green) forming closed
coronal loops and open field lines presumably extending into higher corona and
the solar wind. The red foam represents turbulence. Acceleration probably takes
place in the outflow regions above and below the X-point. Particles
(temporarily) trapped here produce the radiation seen above the closed
loops and particle escaping these regions up and down (blue arrows) are
observed at one AU as SEPs and produce
the nonthermal radiation (mainly at the two foot points; blue ovals),
respectively. (From Liu et al. 2013) {\bf Right:} A similar schematic joining
the flare site field lines to the CME, the shock and beyond (from Lee, 2005).
The rectangles define the
boundary of the acceleration sites and represent the leaky box.}
\label{fig:models}
\end{center}
\end{figure}

The final coefficient in Equation (\ref{KE}), namely
the energy dependence of the escape time, has a more complicated relation to the
transport coefficients and the crossing time $\tcross\equiv L/v$, where
$L=V/{\bar {\cal A}}$ is the average size of the  acceleration site. If
there is a
weak or no trapping of the particles in the acceleration site; i.e. when
magnetic field lines are nearly uniform and we are dealing with the  {\it weak
diffusion limit} with scattering time $\tsc \gg \tcross$ (or mean free path
$\lambda_{\rm sc}\gg L$), then the
particles free stream and escape 
within $\tesc\sim \tcross$. However, the relatively strong HXR
emission observed near the top of many flaring loops (and not from the legs of
the loops; see Masuda et al. 1994;  Petrosian et al. 2002) points to the
presence of some trapping of particles. This can happen
in the {\it strong diffusion limit} with
$\tsc \ll \tcross$, which is one of the 
requirements
for
establishing an  isotropic momentum distribution. Scattering by turbulence is
the most likely source of such trappings.%
\footnote{Scattering by Coulomb collisions cannot be this agent because it is
slower than Coulomb energy loss rate so particles 
lose energy as fast as or faster than they scatter (see Petrosian \& Donaghy
1999). Also as shown by Chen \& Petrosian (2013) Coulomb times are in general
longer than all the other times associated with the transport coefficients.
}
In this case the escape
time is determined primarily by the spatial (and other, see Petrosian 2012)
diffusion
coefficient $\kappa_{ss}$  as%
\footnote{For other less important components of this relation see Petrosian
(2012).}
\beq  %\vspace{-0.2cm}
(\tesc)^{-1}= - {1\over N(t, E)}\int {\cal A}(s)ds {\partial\over
\partial
s}\left(\kappa_{ss}{\partial f(t,s,E)\over
\partial s}\right).
\label{tescDef}
\eeq
If we use the approximation ${\partial f(t,s,E)/\partial s}\sim f/L\sim
N(t,E)/(VL)$
then
the escape
time can be written as 
\beq\label{tausc}
\tesc={\tau^2_{\rm cross}\over \tsc}\,\,\,\,\, {\rm where} \,\,\,\,\, 
\tsc\equiv
3{\kappa_{ss}\over v^2}={3\over 4}\left(\langle{(1-\mu^2)^2\over {\bar
D}_{\mu\mu}}\rangle \right).
\eeq
Combining these two limiting cases we can write $\tesc=\tcross(1+\tcross/\tsc)$.
However, recently Chen \& Petrosian (2013) (CP13), using ``regularized"
inversion technique of Piana et al. (2007), have determined the energy
dependence
of all the coefficients of the kinetic Equation (\ref{KE}) directly and
non-parametrically from \r observations for two flares. These results are
presented as
the crossing time $\tcross$, energy loss time $\tloss=E/{\dot E}_L$, energy 
diffusion
time $\tdiff=E^2/D_{EE}$ and the escapetime $\tesc$. From these and the above
relation between escape and scattering time they obtain the
variation with $E$
of scattering time  $\tsc$ and direct acceleration time $\tac=E/A(E)\sim
p^2/D_{pp}$ for both
shock and stochastic acceleration.
The upshot
of this work is that the deduced energy dependences of the scattering and
acceleration times do not agree
with those calculated by Pryadko \& Petrosian (1997) (PP97) based on
interactions of electrons with  turbulence with varied characteristics (see
Fig. 5 in CP13). This suggests that strong diffusion
may not be the source of the trapping. 

\begin{figure}[!ht]
\begin{center}
\includegraphics[width=15cm]{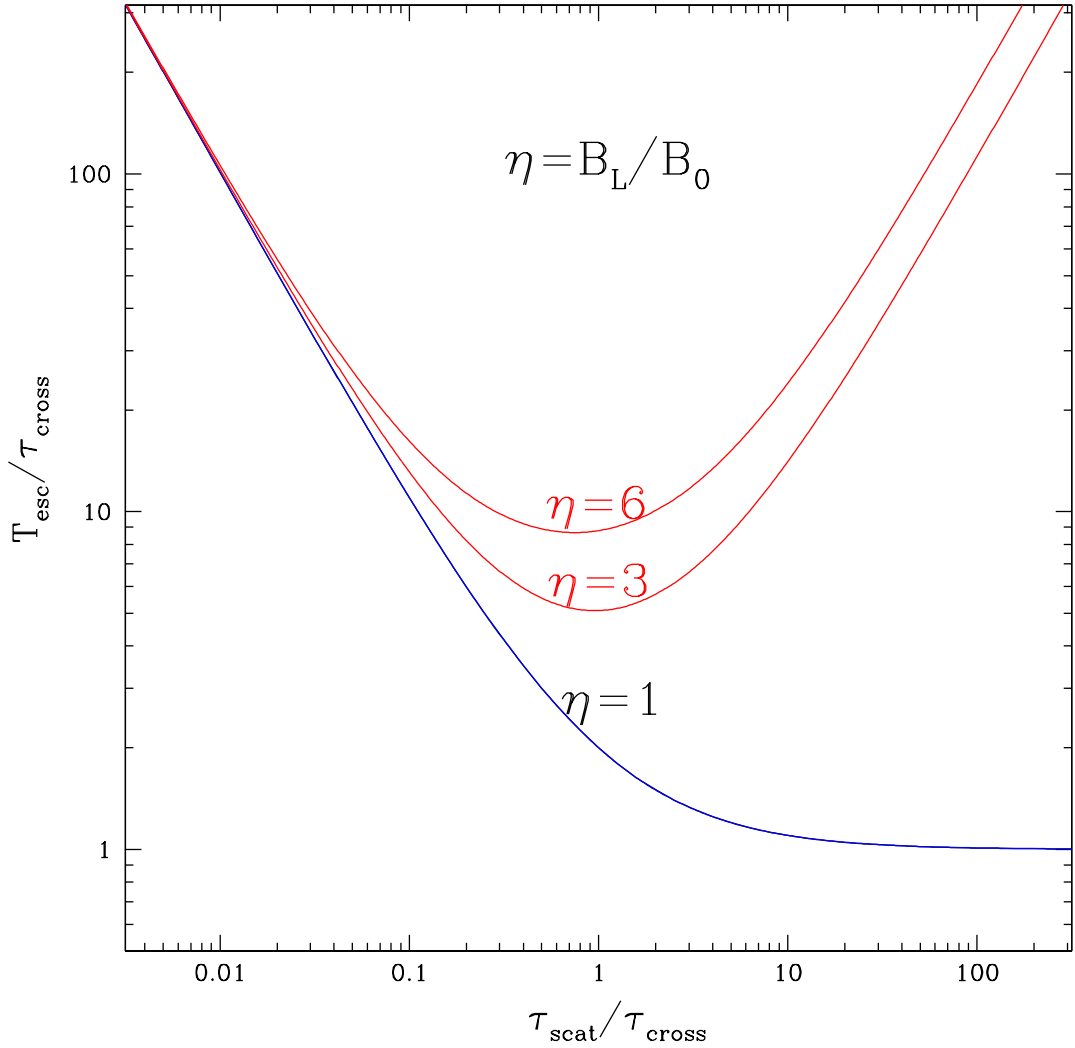}
\caption{Variation of the escape time with scattering time according to Equation
(\ref{tesc}) with $c_1=\eta, c_2=1$ and  $c_3=\ln (\eta)$ for three different
degrees of field
convergence parametrized by $\eta$ the ratio of magnetic field from the center
to the
edges of the acceleration site (size $L$) where particles escape.} 
\label{fig:tesc}
\end{center}
\end{figure}

Transport and trapping of particles in the acceleration site can also be
affected by {\it
magnetic field convergence} from the middle of  the acceleration
site to where they escape at its boundaries. Such a field geometry, shown for
the
toy model 
(Fig. \ref{fig:models} left), seem to be present in some observed loops (Liu et
al. 2013). 
Particles will then bounce between mirror points but  eventually, even in the
weak diffusion limit, they will be scattered  into the loss cone and escape,
so that we expect $\tesc\propto \tsc$. Such a relation will yield a scattering
time  consistent with
the acceleration and energy diffusion times deduced by CP13.
Malyshkin \& Kulsrud (2001) using  some analytic treatment and numerical
simulations, suggest an approximate form for the relation between the
escape and scattering times which can be summarized as

\beq\label{tesc}
\tesc/\tcross=c_1 + c_2(\tcross/\tau_{\rm sc})+c_3(\tsc/\tcross).
\eeq
This equation combines the  strong diffusion case ($\tsc\ll \tcross$) with
$c_2\sim 1$ and $c_1=c_3=0$, and the weak diffusion case ($\tsc\gg \tcross$)
with $c_1\sim 1, c_2=0$. In the  latter case   $c_3\sim 1$ for uniform field
lines and increase with increasing degree of field
convergence. 
Figure \ref{fig:tesc} shows some  examples of the variation of escape time with
the  scattering time (both
in units of crossing time),  where following Malyshkin \& Kulsrud (2001) we have
set $c_1=\eta, c_2=1$ and $c_3=\ln\eta$  with $\eta=B_L/B_0$ the ratio of
magnetic
field strengths at the edge to that at the center of the acceleration site.

In the next section we use the above relations to relate SEP electron spectra
with those
deduced from HXR observations.

\section{SEP AND HXR EMITTING ELECTRONS}
\label{sec:electrons}

The relation between energetic electrons observed near the Earth, with a
velocity
dispersion that indicates their origin from the lower corona region of the Sun,
and those
producing the flare radiation has been subject of several studies. Because
it is by now well
established that type III radio bursts are produced by energetic electrons
moving out from the Sun the first investigations were focused on the
relation between SEP and type III  electrons. Krucker et al. (1999)  using
observations by instruments
on board  the {\it WIND} spacecraft found two classes of events; those with
same release time from the Sun as the type IIIs ({\bf prompt events}) and those
with up to half an
hour delay ({\bf delayed events}) which tend to be harder. Haggetry \& Roelof
(2002) using
higher energy observation by instruments on board the {\it Advanced Composition
Explorer (ACE)} spacecraft find a good association between impulsive SEP
electron
events and other radiative signatures but claim that the electron release time
from the Sun 
is delayed by few minutes relative to the observed radiations (type III,
microwave
and SXRs). They conclude that these electrons are not accelerated at the flare
site but in an outgoing coronal shock. However, the claimed delays have less
than 1.5 sigma significance based on statistical errors alone and there may be
systematic errors such as a longer path than the assumed 1.2 AU or small amount
of scattering.  Maia \& Pick (2004) using observations
by {\it ACE} and the {\it Nancay radioheliograph}, with imaging capability, find
no delay between the electron release and type III times for events that they
classify as
radio-simple (only type III emission) which show  weak association with
CMEs. On the other hand, events classified as radio-complex show variable delays
and are  associated with $>600$
km/s CMEs. They propose ``an acceleration process in the corona, at variable
heights and below the leading edge of the associated" CME.
Somewhat less clear picture is presented by Klein et al. (2005) who, using {\it
Wind} and the {\it Nancay radioheliograph} observations, conclude that a
``combination of time-extended acceleration at
heights $0.5R_\odot$ above the photosphere with the injection of electrons into
a
variety of closed and open magnetic field structures
explains the broad variety of timing shown by the radio observations and the in
situ measurements." 

The main focus of all the above works has been the {\it timing issue}. More
can be learned by investigation of strength and {\it spectral
characteristics}, like hardness or power law index. Although there are some
discrepancies in the above results, they seem to have one  common feature
which is  that the impulsive electron events that appear to
originate from the flare site are weaker than events with more
complex timing relation that are often associated with a CME. It would be of
interest to
see if flare accelerated electrons
show similar trends. This cannot be ascertained by study of the type III or
SXR emissions. Microwave spectra would be useful but need broader  (than 
usually available) frequency
coverage in order to separate optically thin and thick
portions of the spectra. They also depend on the value and geometry of the
magnetic field (see Petrosian  1982). HXR spectra, on the other hand, give a
more direct information about the spectra of the accelerated electrons, and for
a
thick target case this relation is independent of the size of the acceleration
site and
background plasma density. Thus, comparing the spectra of HXRs and  SEP
electrons can shed a greater light on the acceleration process. 

\begin{figure}[!ht]
\begin{center}
\includegraphics[width=8.0cm]{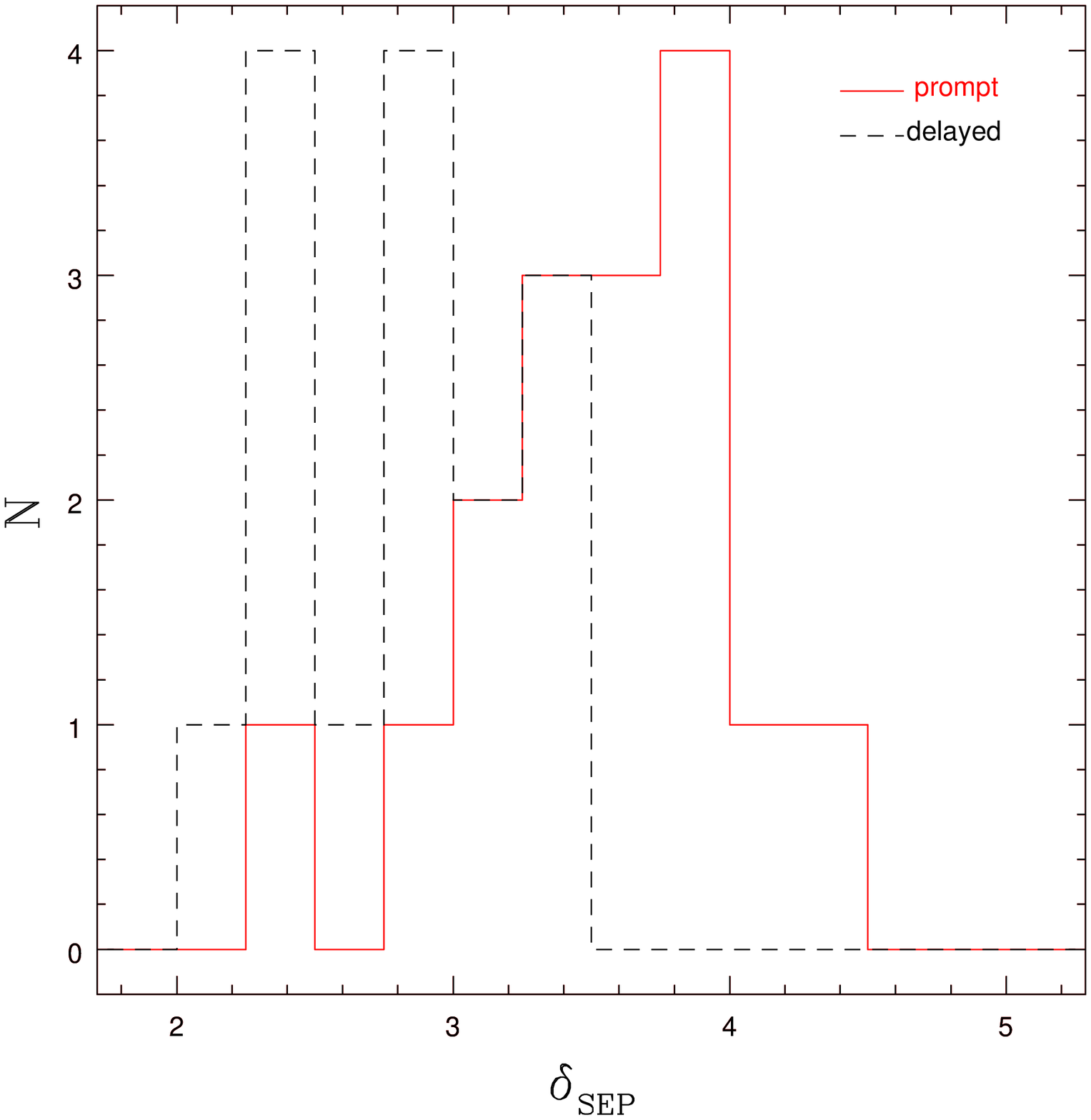}
\includegraphics[width=8.0cm]{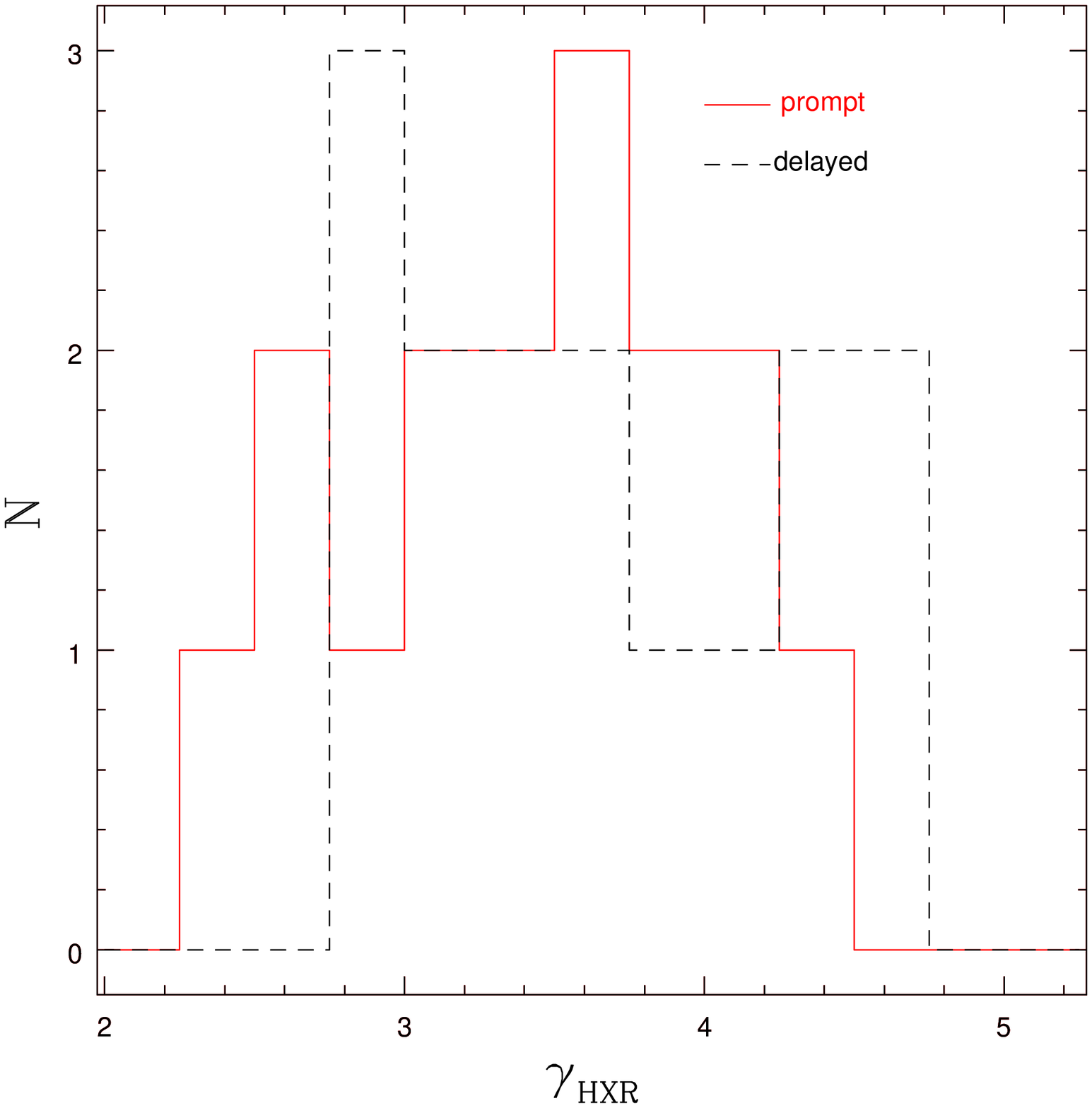}
\caption{Frequency distribution of $\d_{\rm SEP}$ {\bf left} and $\g_{\rm HXR}$
{\bf right} for delayed ( dashed-black) and prompt (solid-red) events, using
the data given in  K07. The average  and standard deviation of the four
distributions are $\d_{\rm SEP}^{\rm
delayed}\sim 2.8 \pm 0.40$,  $\d_{\rm SEP}^{\rm prompt }\sim 3.6 \pm 0.46$), 
$\g_{\rm HXR}^{\rm delayed}\sim 3.7 \pm 0.57$ and $\g_{\rm HXR}^{\rm prompt
}\sim 3.5 \pm 0.55$).} 
\label{fig:hist}
\end{center}
\end{figure}

Krucker et al. (2007) (K07) have carried out such an analysis using \r HXR
spectra and {\it WIND}
electron spectra. They divide their sample into a ``prompt" group, which within
the statistical uncertainties have a release  time in agreement with HXR start
time, and  ``delayed" group. They find a good correlation between the
spectral index $\delta_{\rm SEP}$ of SEP electron flux and spectral index $\g$
of HXR photon flux
above 50 keV with $\delta_{\rm SEP} \sim \gamma_{\rm HXR} +0.1\pm 0.1$ for the
prompt events, but for delayed events they find 
no clear correlation with $\delta_{\rm SEP} < \gamma_{\rm
HXR}$ for majority of such events, i.e, a harder SEP
spectrum. They also find a somewhat weaker correlation between the two
fluxes with HXR observations requiring a significantly larger (by several
hundred) number of radiating electrons than that
observed at one AU. In Figure
\ref{fig:hist} we show the histograms of 
the frequency distribution of the SEP  (left) and HXR (right) indexes for
delayed (dashed-black) and prompt
(solid-red) events. As evident the  the above difference is primarily due to
the fact that SEP spectra of
delayed events are harder than prompt events and
not because their HXR spectra are softer. We find average indexes  $\d_{\rm
SEP}^{\rm
delayed}\sim 2.8 \pm 0.40$ vs $\d_{\rm SEP}^{\rm prompt }\sim 3.6 \pm 0.46$, 
and $\g_{\rm HXR}^{\rm
delayed}\sim 3.7 \pm 0.57$ vs $\g_{\rm HXR}^{\rm prompt }\sim 3.5 \pm 0.55$).
As discussed below, this hardening of the delayed SEP events can be attributed
to the re-acceleration of the flare accelerated  electrons in the CME shock.

\subsection{Accelerated Electron and HXR Spectra}
\label{sec:brem}

The relation between the NTB radiating electron  and the  observed HXR
spectrum 
is relatively simple for a single source. As demonstrated
some times ago this depends whether one deals with a {\it thin target} source,
whereby electrons lose a small fraction of their energy (mostly  by Coulomb
collisions) while radiating, or a {\it thick target} source where they lose all
their
energy (see Lin \& Hudson 1971, Brown, 1972, Petrosian 1973). The spectral
index ${\bar \d}$ of {\bf flux} of electrons are related to the  HXR index
$\g_{\rm
HXR}$ of
bremsstrahlung photon flux as   ${\bar \d}_{\rm thin}=\g_{\rm HXR} -1$ and 
${\bar \delta}_{\rm thick}= \g_{\rm HXR} +1$, respectively.
As pointed out by K07 neither one of these relations agrees with observations
of either the prompt (with $\delta_{\rm SEP}\sim \gamma_{\rm HXR}$) or delayed
events where there is no clear correlations. This may
indicate that  the SEP and radiating electrons have different origins. 

However, the assumption of single radiative source, whether thin or thick is an
over simplification. First, as mentioned above, most flares show emission from
a loop top source and from two (harder) footpoint sources. Second, the spectra
of the  electrons escaping downward, responsible for the radiation,  and 
upwards, observed as SEPs, may be different  from each other, and
most likely different from the spectrum of the accelerated electrons. Thus, the
exact relation between  $\delta_{\rm SEP}$ and $\gamma_{\rm HXR}$ is more
complicated and depends on the details of the acceleration and transport
mechanisms. Let us consider the model for acceleration in the corona shown in
the right panel of Figure \ref{fig:models}. The (red) rectangle represents the
leaky box within which the acceleration mechanism produces a spectrum $N(E)$ of
electrons (integrated over the acceleration region). These electrons are 
responsible for the emission from the loop top source (with  a background
density
of $n_{\rm LT}$) and  yield a  HXR spectrum (photons per s per unit
$\epsilon$) given by 
\begin{equation}
J_{\rm LT}(\epsilon)
=\int_{\epsilon}^\infty n_{\rm LT}vN(E) \sigma(\epsilon, E) dE
=\tau^{-1}_{\rm br}\epsilon^{-1}\int_{\epsilon}^\infty
\beta^{-1}N(E)f(\epsilon/E)dE,
\label{Xrayspec}
\end{equation}
where $\beta=v/c$,  $\tau_{\rm br}\equiv (16/3)r_0^2\alpha nc$ and
$f(\epsilon/E)$ is a slowly
varying function.%
\footnote{For nonrelativistic energies
$f(x)=\ln {1+\sqrt{1+x}\over 1-\sqrt{1+x}}$ and for extreme relativistic regime
$f(x)=(1-x+3x^2/4)[\ln (1.2\epsilon) +\ln {1-x\over x^2}]$ (see Koch \& Motz
1959).}
Here $E$ and
$\epsilon$ are in units of electron rest mass energy $m_ec^2$, $\alpha=1/137$
and $r_0=e^2/m_ec^2$.
This is a {\it thin target} spectrum which in the nonrelativistic regime
($\b=\sqrt{2E}$) and 
for a power law electron spectrum $N(E)=N_0E^{-\delta}$ yields 
\beq
J_{\rm LT}(\epsilon)= {N_0\over \sqrt {2}\tau_{\rm
br}}\epsilon^{-\delta-1/2}I_{\delta -3/2}(0),\,\,\,\, {\rm with} \,\,\,\,
I_n(x)=\int_x^1 t^nf(t)dt,
\label{LTspec}
\eeq
so that the HXR index $\g_{\rm HXR}=\d+1/2$.
Note that in our notation  $\delta$ is the index of  the spectrum accelerated
particles. The index for the {\bf flux} $F=vN$ of electrons  in the
nonrelativistic  regime of interest here will be ${\bar \delta}=\delta-1/2$ and
the HXR index  $\g_{\rm HXR}={\bar \delta} +1$,
the relation mentioned above and used in
K07 and elsewhere.%
\footnote{For the relativistic regime ${\bar \delta}=\delta$ and one gets
$J(\epsilon)\propto
\epsilon^{-\delta}(a\ln \epsilon + b)$; $a$ and $b$ order unity.}

Most of the higher energy HXRs, however, are produced by particles escaping 
downward along the legs
of the loop to the footpoints (see Fig. \ref{fig:models},
left), with a total flux
of $N(E)/T^d_{\rm esc}(E)$,
where they produce {\it thick target} NTB again given by Equation
(\ref{LTspec}) but with  $n_{\rm LT}N(E)$ replaced by $n_{\rm FP}$ times the
effective thick target spectrum (see CP13 and references cited there) 
\begin{equation} 
N_{\rm eff}(E) = {1\over \dot{E}_{\rm L}}
\int_E^{\infty} {N(E^\prime)\over T^d_{\rm esc}(E^\prime)} dE^\prime,
\label{Neff}
\end{equation}
where  the energy loss rate at the foot points, dominated by Coulomb
collision in nonrelativistic energies, is   
${\dot E}_{\rm L}=1/(\beta\tcoul)$; with  $\tcoul= 1/(4 \pi r_0^2 nc
\ln
\Lambda)$ and Coulomb logarithm $\ln \Lambda \sim 20$. The total bremsstrahlung
flux will be equal to the sum of the loop top and footpoint emissions. In the
Appendix A we describe an inversion method where one can obtain the  accelerated
and effective spectra from the total spectrum for a given energy dependence of
the escape time. However, for our purpose here dealing with spectra and fluxes
above
50 keV the loop top emission can be ignored so we will consider the thick
target footpoint spectra. For the limited range of available observation
 K07 fit the HXR spectra with  power laws and give the indexes. As evident from
Equation (\ref{LTspec}) (and Eq. (\ref{FPspec}) below), this requires a
power-law accelerated electron
spectrum. If we  assume  a power law  energy dependence of downward escape
time, $T^d_{\rm esc}(E)=T^d_{\rm esc,0}(E/E_0)^{\alpha_d}$, (which seems to be
a good approximation; see CP13), then it is straight
forward to show
that 
\beq
J_{\rm FP}(\epsilon)= {N_0\over \d +\a_d-1}{\tcoul\over T^d_{\rm esc,0}\tau_{\rm
br}}\epsilon^{-\d-\a_d+1}I_{\delta +\a_d-3}(0),
\label{FPspec}
\eeq
so that the HXR spectral index is
\beq
\g_{\rm HXR}=\d +\a_d -1.
\label{gHXR}
\eeq
Note that for energy independent escape time ($\a_d=0$) this gives the standard
thick target relation.
This relation is valid both for prompt and delayed events. In the next two
sections we relate these HXR spectral indexes to the SEP electron index.

\subsection{Prompt Events}
\label{sec:prompt}

Some of the accelerated electrons, those on open field line, escape upward and
out
of solar atmosphere with an escape time $T^u_{\rm esc}(E)$, and  are observed as
SEP electrons. If they are not scattered or have not suffered  any energy
loss or gain, which is believed to be the case for the prompt events, then the
observed SEP flux spectrum will be proportional to $N(E)/T^u_{\rm esc}(E)$.
Assuming a power law escape time (which again is a good approximation for small
range of energies 50 to $<500$ keV), $T^u_{\rm esc}(E)\propto  E^{\a_u}$,
the SEP electrons will have an spectral index $\d_{\rm SEP}=\d
-\a_u$. Combining this with Equation (\ref{gHXR}) we get 
 \beq
\d_{\rm SEP}=\g_{\rm HXR}+1+\a_u -\a_d. 
\label{dSEP}
\eeq
Thus, it is clear that  {\it the energy dependences of the escape times down to
the
photosphere and up  into the solar wind play crucial roles in
the relation between the SEP electron index and the HXR index.} Consequently,
the
observed indexes can be used to determine the escape time indexes $\a_u$ and
$a_d$, and constrain the models. For example, the usual assumption used  in the 
standard thick target model that the escape times
are independent of energy (i.e. $\a_u=\a_d=0$), lead to $\d_{\rm SEP}=\g_{\rm
HXR}+1$, which can be ruled by the observed linear regression relation $\d_{\rm
SEP}=\g_{\rm HXR}+0.1 \pm 0.1$ given in K07. 
$\pm 0.1$ is the error on the slope and not the intercept. As evident from 
Figure {\ref {fig:hist} the dispersions (given in the caption) of measured
spectra are larger. In what follows we use this larger error bar of $\pm 0.3$,
which is also comparable to errors of individual index measurements. Thus, we
estimate that thick and thin target models are ruled at the 3$\sigma$ level.

In the {\bf strong diffusion case} particle escape up and down is determined
by the scattering and crossing time in the acceleration site which would
give the same energy dependence for both escape times, $\a_d=a_u$. This also
leads to a thick-target type relation and  can be ruled at
about  the same level. 

In the {\bf weak diffusion limit}, on the other hand,  
for the electrons escaping upward along {\it diverging} field lines we are
in the $\eta=1$, $\tcross<\tsc$ branch of Figure \ref{fig:tesc} so
that $\tesc\sim \tcross=L/(c\b)
\propto E^{-0.5}$ at nonrelativistic energies and approaches a constant $L/c$
in the extreme relativistic regime. More exactly  $\a_u=-d\ln \beta/d\ln
E=1/[(E+1)(E+2)]$. In the energy range of interest, 50 to 200 keV this
changes from $-0.43$ to $-0.3$. In what follows we will use $\a_u=-0.4$. From
Equation (\ref{dSEP}) we
then get the average value ${\bar \a_d}=0.6\pm 0.30$. This positive index
is in agreement with the energy dependence obtained by CP13 for two flares.
As  derived in CP13 the acceleration time also
increases with increasing energy. Because, in general we expect similar energy
dependence for the scattering and acceleration times (see the discussion
in \S 2 and  PP97), the above result implies that we are again in the weak
diffusion limit but on the branches with $\eta >1$. To determine the value of
$\eta$ we need the absolute values of the escape and scattering time. The
latter depends on the intensity and spectrum of the turbulence (see the
discussion below) which is unknown or at least cannot be derived from the data
under consideration here. A corollary of this result is that from the
analysis of this  kind of data we can constrain the energy dependences of the
scattering and acceleration times.

Figure \ref{fig:alphad} shows the frequency distribution of $\a_d$ we
obtain from K07 observations. 
As also shown in this figure, the two flares for which CP13
derive the values of  this index are in excellent agreement with the
K07 observations. 
{\it We conclude, therefore, that the observations presented in K07 for
the spectra of prompt events are consistent with the idea that the trapping of
the electrons in the loop top region is caused by converging field lines from
loop top to footpoint regions of the reconnecting flare loop and that we are
in the weak diffusion limit.}
\begin{figure}[!ht]
\begin{center}
\includegraphics[width=15cm]{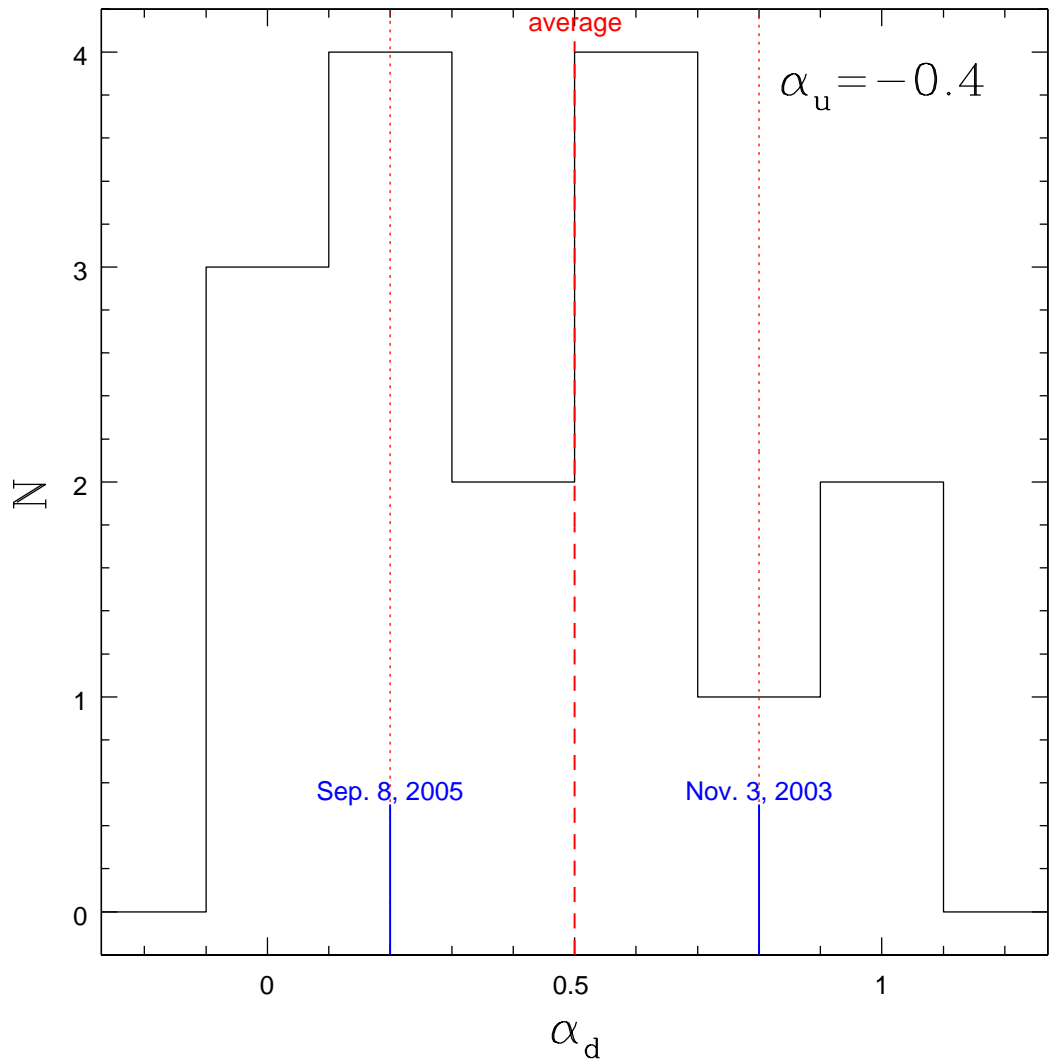}
\caption{Frequency distribution of $\a_d$, the power law index of energy
dependence of the escape time downwards, based on Equation (\ref{dSEP}) and
the observed
values of indexes of prompt events, in the weak diffusion limit with $\a_u=-0.4$
(i.e. free streaming of accelerated electrons upward from the acceleration site
along diverging open field lines). The dashed and dotted (red) vertical lines
give the average and one sigma range of the observations. The short (blue)
vertical lines show the value of $a_d$ obtained by CP13 for two flares using
an inversion method.} 
\label{fig:alphad}
\end{center}
\end{figure}

Next we consider the {\bf relative numbers of observed SEP electrons and the
number
of radiating electrons} deduced from HXR fluxes. We first define the total
spectrum  of
accelerated electron  $N_{\rm tot}(E)=\int N(E,t)dt=N(E)\Delta T$,
where $N(E)$ is the average (or  steady state) spectrum of electrons and $\Delta
T$ is the flare duration. We  express the time integrated number of SEP and
HXR-producing electrons in terms of these two quantities. The expected flux of
SEP electrons at Earth (distance $d$), after
correcting for the velocity dispersion,  $F_{\rm SEP}= (N(E)/T_{\rm
esc}^u)/(\Omega d^2)$, where $\Omega$ is the sterradians within which the
up-escaping
electrons are
directed. If one assumes that the spatial distribution of SEPs is
isotropic,%
\footnote{Note that the velocity distribution of on diverging field lines will
be peaked around the direction of the field lines. But since we are dealing
with pitch angle integrated quantities this does not affect the results
presented here.}
then the deduced number of SEP electrons above some energy $E_0$ (here 50 keV)
will be given by 
\begin{equation}
N_{\rm SEP}=\left({4\pi \over \Omega}\right) \left({c\Delta T\over
L_u}\right) \int_{E_0}^\infty 
\b(E)N(E)dE=\left({4\pi \over \Omega}\right) \left({c\Delta T\over
L_u}\right)N(E_0){E_0^{1-\a_u}\over \g_{\rm
HXR} -\a_d}
\label{NSEP}
\end{equation}
where we have defined $N(E_0)=N(E)(E/E_0)^{\d}$ and $L_u=V/{\cal A}_u$
(with
${\cal A}_u$ the cross section area of the bundle of field lines over which the
particles escape to the Earth), and  used the weak diffusion
escape time (${\a_u=-d\ln \b/\d\ln E}$)  and Equation (\ref{dSEP}). 

The total (volume and time integrated) number of HXR electrons are obtained from
the total number of observed HXRs
with energies $\epsilon\geq E_0$, which following Equation (\ref{LTspec}) can be
written  as
\beq\label{TotHXR}
J(\epsilon>E_0)=\Delta T n{\cal A}_d\int_{E_0}^\infty d\epsilon
\int_{\epsilon}^\infty F_{\rm eff}(E)) \sigma(\epsilon, E) dE=\Delta T
n{\cal A}_d\int_{E_0}^\infty  F_{\rm eff}(E)) dE\int_{E_0}^E
\sigma(\epsilon, E)d\e,
\eeq
where $n$ is the background density, ${\cal A}_d$ is the area and
$F_{\rm eff}(E)=vN_{\rm eff}(E)/V$ is the effective mean (or the steady state)
flux
of
electrons. Inversion of this equation can give the total number of HXR
emitting electrons 
$N_{\rm HXR}=(\Delta Tc/L_d)\int_{E_0}^\infty \b N_{\rm eff}(E)dE$, with
$L_d=V/{\cal A}_d$. As
mentioned above the thick target emission is independent of the background
density so that in the above relations one can use a fiducial value for
the density because  the energy loss
rate is  ${\dot E}_{\rm L}=4\pi r_0^2\ln \Lambda cn/\b$ in the
denominator of Equation (\ref{Neff}) is proportional to $n$ (for further
clarification see CP13). Now substitution of the
integral in (\ref{Neff})  in the above expression (and after
exchanging the order of the integrations over $E^\prime$ and  $E$) we obtain 
\beq\label{NHXR}
N_{\rm HXR}={\Delta Tc\over L_d}{\tcoul(E_0)\over
T^d_{\rm esc}(E_0)}{E_0^{\a_d -1}\over \beta(E_0)}\int_{E_0}^\infty
{N(E^\prime)\over
E^{\prime,\a_d}}dE^\prime\int_{E_0}^{E^\prime} \b^2dE,
\eeq
where we used the power law approximation for the escape time
[$\tesc(E)=\tesc(E_0)(E/E_0)^{\a_d}$]. In the non nonrelativistic
limit, $\b^2\sim 2E$,  Equations (\ref{NSEP}) and (\ref{NHXR}) give the
ratio 
\beq
R_N={N_{\rm HXR}\over N_{\rm SEP}}= \left({\Omega\over 4\pi}\right)
\left({L_u\over L_d}\right)\left({\tcoul(E_0)\over
T^d_{\rm esc}(E_0)}\right)\left({\g_{\rm
HXR} -\a_d \over  (\g_{\rm HXR}+\a_u)(\g_{\rm HXR}+\a_u-2)}\right).
\label{ratio}
\eeq
The primary term here is the ratio of the the Coulomb to escape time which for
the two flares measured by CP13 is about 10 but increases with decreasing
value of the loop top density.. The first term in the ratio depends on the
spatial anisotropy of the SEP electrons which is is not well-known and could be
close
to unity. The last term for $\a_u=-0.4$ and for the observed range of 
$\a_d$ shown in Figure \ref{fig:alphad} varies from about 0.5 to 10. The final
term $L_u/L_d={\cal A}_d/{\cal A}_u$ depends at what
point the acceleration stops below and above the X-point reconnection. This
ratio is most
likely smaller than 1. The factor of 10 derived here is smaller than the
observed ratios of about several hundred, indicating perhaps that densities are
about 10 times lower than $n_{\rm
LT}\sim (2 -5)\times 10^{10} {\rm cm}^{-3}$ used in CP13, or that a smaller
fraction of upward escaping electrons are on field lines with good
connection to the Earth.

\subsection{Delayed Events and Re-acceleration}
\label{sec:delayed}

As shown in Figure \ref{fig:hist}, the  SEP spectra of delayed events on
average are harder than those of the prompt events and also  harder than
that of the associated
HXR producing electrons. Moreover, the  correlation between $\d_{\rm SEP}$ and 
$\g_{\rm HXR}$ is not as strong as that in the prompt events. The
temporal relation of SEPs and HXRs and the weak correlation between their
indexes (see  bottom panel of  Fig. 3 in K07) indicate some  connection
between the SEPs and HXRs in the delayed events as well. As shown in Figure
\ref{fig:r} for several
events $\g_{\rm HXR}\sim \d_{\rm SEP} +1.5$ indicating that $\a_d=\a_u+2.5>2$.
In the weak diffusion limit this will require a turbulence spectrum that is much
flatter than kolmogorov and in the strong  diffusion limit a very steep
spectrum. Thus, this behavior may be an
indication of presence of a  different acceleration process. Since
delayed-gradual events are more likely to be associated with Type II radio
bursts and/or  CMEs, then it is natural to assume that the SEPs in these events
are accelerated in the CME shock. 

The nature and origin of the seed
particles in
shock acceleration in general, and for production of  SEPs in CME
shocks in particular, is not well known. It is sometimes assumed to be a
preexisting nonthermal component in the
solar wind with a {\it kappa} type distribution, or particles trapped
behind an earlier (slower) CME (see e.g Parker \& Zank 2012; Tylka \& Lee 2006
and references therein).  In these and most other scenarios of SEP
acceleration in a CME shock the seed particles come from the upstream
region of the shock. Here we consider an alternative model where seed
particles come from the downstream region. Because of the strong temporal
relation, the weak
spectral correlation and
consistently harder spectra of the delayed events we are lead to consider a
scenario where the SEP spectra result from re-acceleration of flare site
electrons. This re-acceleration is the cause of the harder  SEP than HXR
producing electron spectra  in these events.%
\footnote{Note that a shock, or for that matter any other mechanism, does
not always result in a harder spectrum. A seed population with a hard spectrum
will be modified slightly by a mechanism with weak acceleration rate; see e.g
Forman, Webb \&  Oxford (1981). As we will show below in our case dealing with
relatively soft (flare accelerated) seed particles, a reasonable rate of
acceleration can lead to a harder spectrum.}
An important assumption here is that the CME  launch and acceleration of the
electrons at the reconnection are almost simultaneous, and/or the upward
escaping accelerated electrons are trapped by turbulence (or other means) in
the downstream region of the shock. As described in \S 3.2 we need
only a small fraction of these electrons to be trapped and re-accelerated, which
renders this assumption reasonable.

\begin{figure}[!ht]
\begin{center}
\includegraphics[width=15cm]{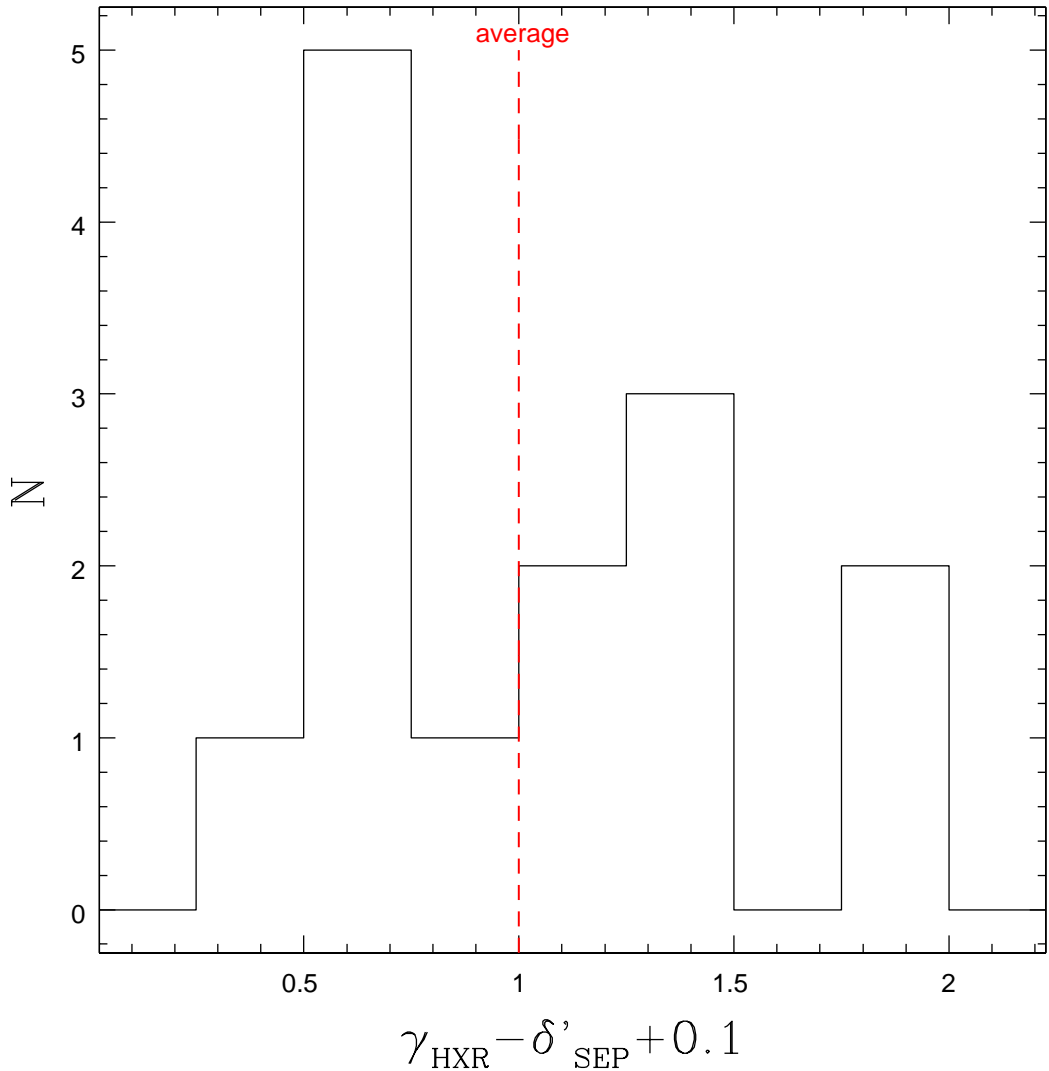}
\caption{Frequency distribution of degree of hardening observed in delayed
SEPs compared to prompt events. We assume that the electrons escaping the
acceleration site have an spectral index $\d_{\rm ESP}= \g_{\rm HXR}+0.1$, thus
this represent the difference in the index of the electrons escaping the flare
site and the
index of those escaping the CME environment after re-acceleration there. This
change of index is related to the index $r\equiv d\log {\cal R}(E/d\log E$.} 
\label{fig:r}
\end{center}
\end{figure}

In order to treat the re-acceleration we must now consider the solution of the
kinetic
Equation (\ref{KE}) in the leaky box depicted on the right panel of Figure
\ref{fig:models}, where now the source term is the spectrum of particles
escaping
upward from the Sun; ${\dot Q}^\prime =N(E)/T_{\rm esc}^u(E)$, which we assume
to be a
power law with index $\d_{\rm SEP}\sim \g_{\rm HXR}+0.1$ as in the
prompt events.%
\footnote{In what follows we will use primed quantities for the re-acceleration
process.}
The background plasma conditions in the CME shock is quite different than
that in the flare loop. Both magnetic field and density are lower and decrease
with increasing distance from the Sun. 
Gopalswamy \& Yashiro  (2011) using the observed sizes and radii of the CME and
the
shock, and theoretical model of Russell \& Mulligan (2002), derive
an Alfv\'en Mach number
${\cal
M}_A\sim 2$ and Alfv\'en velocity $v_A\sim 500$ km/s. From this they
estimate
the plasma density and magnetic field values and variations with distance. At a
distance of 2 to 3  solar radii (from the center of the Sun) they  give  $B\sim
0.1$ G and $n\sim
10^5$ cm$^{-3}$. This implies a low energy loss rate or  energy loss times 
$\tloss > 10$ days so that the loss
term in Equation (\ref{KE}) can be ignored. Let us first assume that the
acceleration
by the CME driven shock is the  dominant mechanism here in which case we can
ignore the energy diffusion term and obtain  the kinetic equation for the steady
state case as
\beq\label{KE'}
d(A^\prime N^\prime)/dE+N^\prime/T^\prime_{\rm esc}={\dot Q}^\prime.
\eeq
If we define $d\eta=dE/(T^\prime_{\rm esc}A^\prime)$ we obtain  the formal
solution for the flux of escaping electrons
\beq\label{solution}
F^\prime(E)={N^\prime(E)\over T^\prime_{\rm esc}(E)}={{\cal
R}(E^\prime)\over E}\int^E_{E_0} e^{\eta(E^\prime)-\eta(E)}{\dot
Q}^\prime(E^\prime
)dE^\prime,
\eeq
where $E_0$ is a fiducial value of energy below which the approximations used
breakdown and we have defined 
\beq\label{eta}
{\cal R}(E)={E\over A^\prime T^\prime_{\rm esc}}={\tau^\prime_{\rm ac}(E)\over
T^\prime_{\rm esc}(E)}\,\,\,\,\,\, {\rm and }\,\,\,\,\,\,
\eta(E^\prime)-\eta(E)=\int^{E^\prime}_E {\cal
R}(x)d\ln x.
\eeq
As shown in Appendix B a slightly different version of this solution provides a
good approximation for stochastic acceleration scenario too.

The solution of this integral is governed by the energy dependence of the  ratio
of acceleration time to escape times ${\cal R}(E)$. If
for all relevant energies  ${\cal R}(E)\gg 1$ the electron escape at
a faster rate than they are accelerated. Thus, there
will be little re-acceleration and the escaping electrons from the CME
environment will have the same
spectrum   as the injected electrons escaping the flare site;
$F^\prime(E)={\dot Q}^\prime(E)$, and have the same spectral
index; $\d^\prime_{\rm ESP}=\d_{\rm ESP}$. This result also can be derived from
Equation (\ref{solution}) by noting that in this case $\eta(E)\sim {\cal R} \gg
1$ so that most of the contribution to the integral comes from the upper limit
of the integrand; $\int^E_{E_0} ...\sim {\dot
Q}^\prime dE/d\eta=A^\prime(E) T^\prime_{\rm esc}(E){\dot Q}^\prime(E)$.
In the
opposite extreme
case,
$\eta(E)\sim {\cal R}(E)\ll 1$, particles undergo considerable acceleration
before they escape. In this case  the exponential terms in Equation
(\ref{solution}) are about one, and for $\d_{\rm SEP}>1$, which is the case
here, most of the contribution to the integral comes from the (constant) lower
bound of the integral, so that the integral is  essentially independent of
the energy spectrum of the injected flux and the re-accelerated flux
$F^\prime(E)\propto
E^{r-1}$, where $r\equiv d\log {\cal R}/d\log E$. As shown in  Appendix C this
change of spectrum
with ${\cal R}$ can be seen clearly for the special case of $r=0$, i.e.
when  $\tac$ and $\tesc$ have similar energy dependences, where one can
get
an analytic solution. These behaviors is also evident
in Figure \ref{fig:sol}, where we present re-accelerated fluxes
($F^\prime(E)/[{\dot Q}^{\prime}(E_0){\cal R}_0]$) from numerical evaluations of
Equation (\ref{solution}) for two values of the spectral index of injected
electron (from the flare site) and four values of the ratio ${\cal R}(E_0)$, at
a fiducial energy  $E_0\sim 50$ keV, and two values of its exponent $r$.%
\footnote{In order to avoid divergence at low energies we use the injected
spectrum 
${\dot Q}=[1+(E/E_c)^\d]^{-1}$.}

As evident for large ${\cal R}$ and at higher energies we get spectra
as steep as the injected spectra but at lower energies and lower values of
${\cal R}$ we  get  harder
spectra with index   $\d^\prime_{\rm SEP}\sim r-1$, independent of injected
spectrum. Therefore, in order to produce the harder spectra (change of index by
1 or 2) seen in the delayed events, we
require a value of ${\cal R}$ slightly less than one and an index  $r\sim 1$.
Higher
values of this index can result in an index changes ranging from 3 to 4  in
the energy range $E_0/2<E<5E_0$.

{\it A more important result is that  this kind of analysis can
shed light on the characteristics of the re-acceleration mechanism.}
Below we give a brief description of the expected values of these
characteristics.

\begin{figure}[!ht]
\begin{center}
\includegraphics[width=8.0cm]{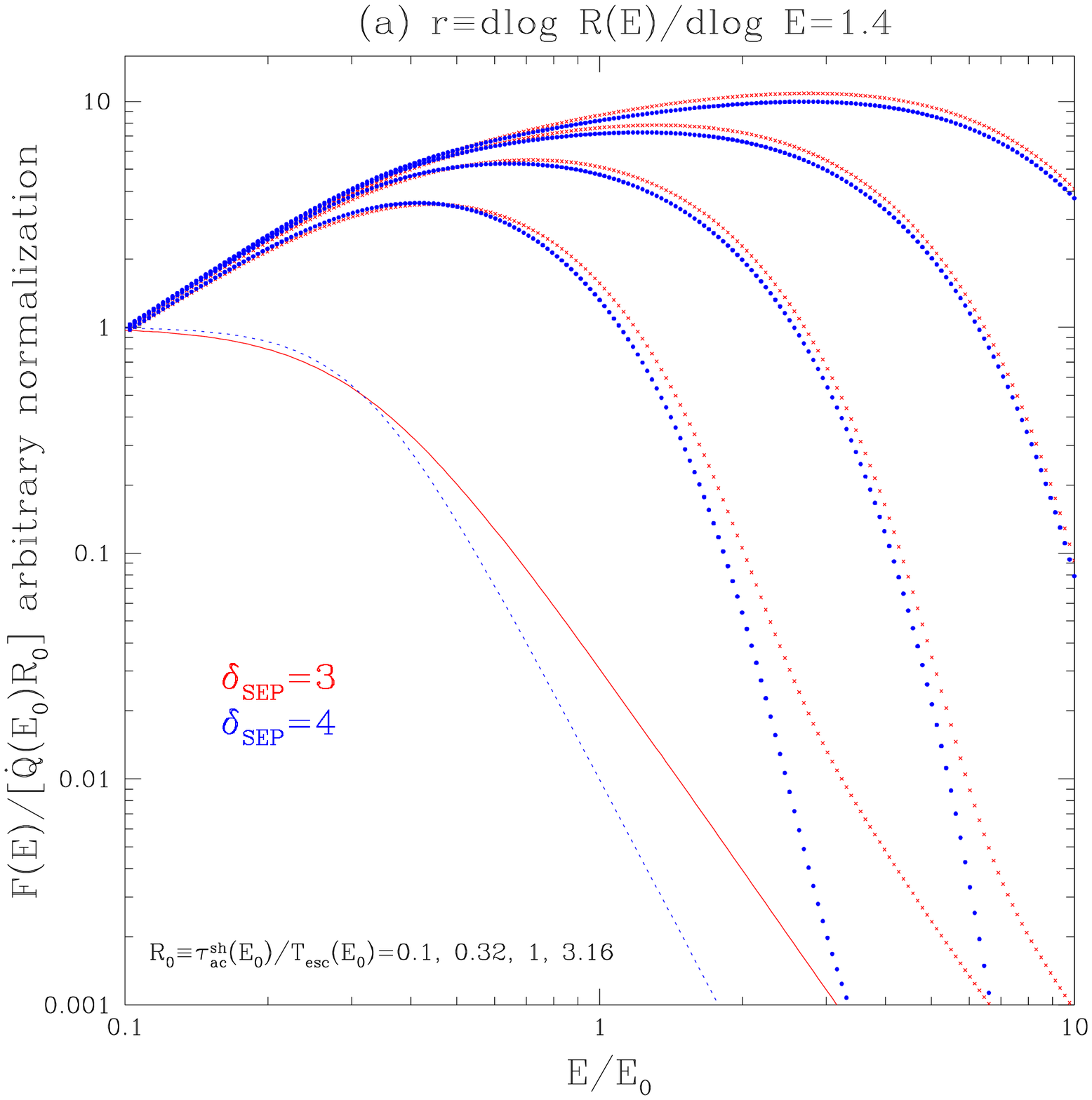}
\includegraphics[width=8.0cm]{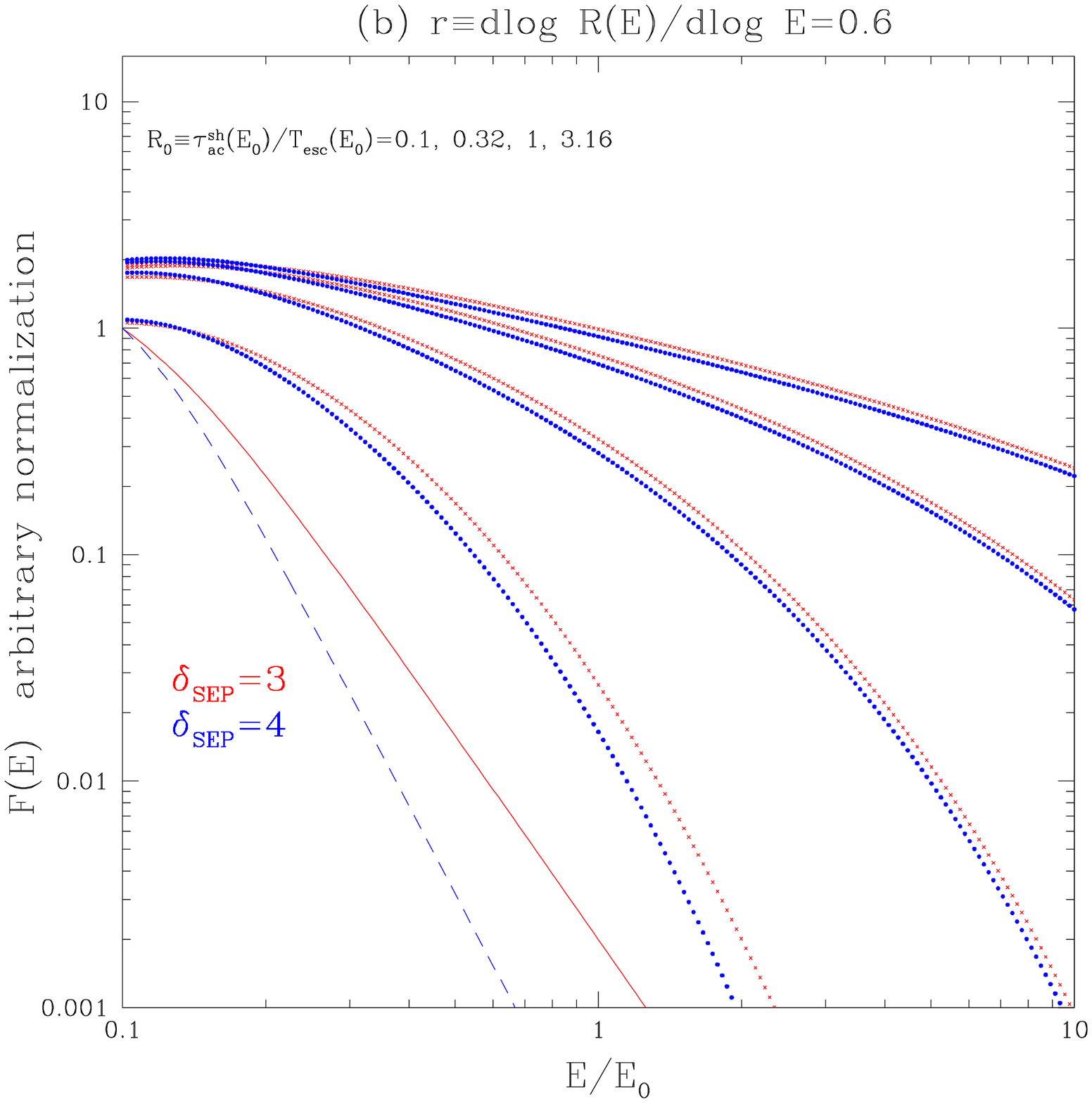}
\caption{Flux of re-accelerated spectra  of electrons
injected by the coronal flare into the CME environment with two spectral
indexes $\d_{\rm SEP} =3$ (red-crosses) and $\d_{\rm SEP} =4$ (blue-dots), and
for two values of the index $r\equiv d\log {\cal R}(E)/d\log E$. Note  that
spectra change rapidly from no re-acceleration to highly hardened spectra for
decreasing values of the ratio {\cal R} from bottom to top. The two curves
(solid-blue, red-dotted) show the assumed injected spectra ${\dot
Q}(E)=[1+(E/E_c)^\d]^{-1}$, with $E_c\sim E_0/3$, chosen to avoid divergence at
low energies, but to have the simple power law form in the relevant energy range
$E>E_0/2$.} 
\label{fig:sol}
\end{center}
\end{figure}

\subsection{Requirements of Re-acceleration}
\label{sec:acc}

For
re-acceleration to be significant we need a acceleration time somewhat
shorter than the escape time, but not very much shorter because the spectral
hardening will be too severe. In the CME environment, with relatively low
magnetization and Alfv\'en velocity, the scattering time is expected to be much
shorter than the acceleration time so that $\tesc\gg \tsc$. In the turbulent
upstream and downstream of the shock this implies we are dealing with the strong
diffusion case which is
required for repeated passage of the electrons across the CME
shock if acceleration is indeed by the shock. Short acceleration time is  also
required if the
acceleration is  directly (and stochastically) by the turbulence. 
From  Equations
(\ref{accrates}) and (\ref{tausc}) we obtain the stochastic and shock
acceleration times $\tau^{\rm SA}_{\rm ac}=E/A^\prime(E)=\zeta^\prime\tau_{\rm
ac}$ (see Appendix B for definition of $\zeta^\prime$) and
$\tau^{\rm sh}_{\rm
ac}=(3/4)\zeta^{-1}(v/u_{\rm sh})^2\tsc$ from which we
derive 
\beq\label{tratio}
{\cal R}_{\rm sh}(E)={3\over 4\zeta}\left({v\over u_{\rm
sh}}\right)^2\left({\tsc\over
\tcross}\right)^2\,\,\,\,\, {\rm and}\,\,\,\,\, 
{\cal R}_{\rm SA}(E)={\zeta^\prime\tau_{\rm ac}\tsc\over \tau_{\rm cross}^2}.
\eeq
Thus, the degree of hardening due to re-acceleration is determined by the
acceleration and scattering times which depend on the momentum (or energy) and
pitch angle diffusion coefficients for stochastic and shock acceleration,
respectively. These in turn are determined by the physical condition of the
plasma (primarily the Alfv\'en velocity $v_A=c\beta_A$), and spectral
charcteristics of the turbulence (primarily the spectral index $q$ and the range
of
$k_{\rm min}, k_{\rm max}$ of the $k$ vector) in the re-acceleration site. For
extreme relativistic electrons, interacting mainly with Alfv\'enic turbulence,
these
relations are simple; $\tau_{\rm ac}^{\rm SA}\sim \tsc/\beta_A^2 \propto \tau_p
E^{2-q}$,
where the characteristic rate $\tau_p^{-1}\sim \Omega_c f_{\rm turb} (ck_{\rm
min}/\Omega_c)^{q-1}$, with $\Omega_c$ the gyrofrequency and $f_{\rm turb}\sim
(\delta B/B)^2$ the fraction of the magnetic energy in turbulence. But at the
sub- and semi-relativistic energies of interest here electrons interact with
other modes
of the turbulence and the relation between the two time scales and their energy
dependences vary significantly with the values of $q$ and $\beta_A$. Examples of
these were presented first in Figs. 12
and 13 of PP97 (for a combined version see  Fig. 5 of
CP13). Note that PP97, instead of the Alfv\'en velocity,  use the
ratio of plasma to gyro frequencies $\alpha=\omega_p/\Omega_c=1/(43\beta_A)$.
Moreover, the rates calculated in PP97 are for parallel propagating waves
including all cold plasma modes. Subsequent papers Pryadko \& Petrosian 1998,
1999 deal with perpendicular propagating waves and hot plasma effects and
obtain similar results. In what follows we will use the
results for parallel propagating waves.
As mentioned above in typical CME environment at a distances of $\sim
2R_\odot$ the Alfv\'en velocity is about
500 km/s (corresponding to $\alpha\sim 14$), 
and although the plasma
density and magnetic field decrease rapidly with  distance from the Sun,  the
Alfv\'en velocity decreases only gradually.
Fitting the curves in PP97 for $\alpha=10$ and $q=5/3$, in the energy
range 50-200 keV,  we obtain $\tac/\tau_p\sim 20 E_{50}^{\sim 0.7}$
and
$\tsc/\tau_p\sim 1.0 E_{50}^{\sim 0.0}$, where $E_{50}\equiv (E/50 \,{\rm
keV})$. From these we get the timescales
\beq\label{tacfit}
{\tau_{\rm ac}^{\rm sh}(E)\over \tau_p}={0.13\over \zeta\beta_{\rm
sh}^2}E_{50}^{0.8},\,\,\,\,\ {\tau_{\rm ac}^{\rm SA}(E)\over
\tau_p}=20\zeta^{\prime}E_{50}^{0.7}\,\,\,\,\,{\rm and} \,\,\,\,\,
{T^{\prime}_{\rm esc}(E)\over \tau_p}= 5.9 \left({L\over
c\tau_p}\right)^2E_{50}^{-0.8},
\eeq
where we have used the approximation
$\beta=0.4E_{50}^{0.4}$.
These give  similar ${\cal R}$'s and hardening indexes of $r_{\rm sh}\sim 1.6$
and $r_{\rm SA}\sim 1.5$, which as shown in Figure \ref{fig:sol}a, results in
significant
hardening but somewhat curved spectra.
Carrying out this exercise for $q=3$ we get index values of 0.7 and 0.4,
which will give less curved spectra  similar to those in Figure
\ref{fig:sol}b.

{\it We therefore  conclude that the properties of the observed sample of
delayed events as a whole is consistent with primary acceleration in the corona
and re-acceleration  in the CME environment.} 

To compare {\bf stochastic vs shock acceleration} further we calculate the
ratio of their rates or time scales
\beq
R_{\rm ac}={\tau_{\rm ac}^{\rm sh}\over \tau_{\rm ac}^{\rm SA}}={3\over
4\zeta\zeta^\prime\b_{\rm sh}^2}\left({\tsc\over \tac}\right)\sim
58\times E_{50}^{0.1}
\eeq
where we have used  a relatively high  Alfv\'en velocity of 1000 km/s, and set
$\zeta=10$ and $\zeta^{\prime}=1$. In spite
of the relatively high Alfv\'en velocity {\it it appears that the conditions in
the CME environment of delayed events
favor re-acceleration of semi-relativistic electrons directly  by turbulence
rather than by the CME shock}. It should be noted that the above
calculations are based on acceleration and scattering by parallel propagating
waves. However, as mentioned above we expect similar results for obliquely
propagating waves.

\section{ACCELERATION OF He IONS}
\label{sec:He}

We now briefly address similar dichotomy that is observed in the
charcteristics of SEP ions. We will focus on
relative spectra and abundances of $^3$He and $^4$He.  Accelerated  ions
directed to the foot points of flaring loops produce $\gamma$-ray lines (and
pions that decay into $> 70$ MeV gamma-rays). However, these are dominated by
protons and in most flares cannot be used to determine the contributions of
other ions. Thus, in the absence of detectable
radiative signature for these ions we have only SEP spectral and temporal
characteristics at our disposal. As mentioned in \S \ref{intro} there are
differences similar to those found for electrons in the characteristics of ions;
with shorter, weaker events (called ``impulsive') having softer spectra and
higher enhancements of $^3$He and heavy ions compared to, stronger, long
duration (called ``gradual")
events which are associated with fast CMEs (see e.g. Reames et al. 2014).%
\footnote{It should be emphasized that there is no evidence for a 
impulsive-gradual or high-low enhancement bi-modality. As shown by Ho et al.
(2005) (see also Petrosian et al. 2009) there is a broad continuum  of $^3$He to
$^4$He fluence ratio
that decreases with increasing $^4$He fluence.}

Here we explore the
possibility of similar explanation with coronal acceleration of the impulsive
SEPs and re-acceleration
in the CME environment in stronger longer lasting events.  Stochastic
acceleration by
turbulence has been the working hypothesis for gamma-ray producing ions (see,
e.g.~Ramaty \& Murphy 1987) as well as some SEPs (see e.g.  Mason et
al.~1986; Mazur et al. 1992; Miller 2003;
Ng \& Reames 1994), but  CMEs and the resultant shocks are believed to play
important roles for gradual SEPs (Zank 2012).

One of the most vexing problems of SEPs has been the  extreme enhancement of
$^3$He (e.g.~Nitta et al.~2008).
It was recognized early that the plasma wave-particle interactions and
the unusual charge to mass ratio of $^3$He
could be the cause of such enhancements (see e.g.~Fisk 1978 and other early
works referred to
in Petrosian et al.~2009).
More recently, and with a more detailed treatment of the plasma wave
interactions with $^3$He and $^4$He ions,  
Liu et al. (2004 and  2006) demonstrated that stochastic acceleration by
turbulence can quantitatively explain both the extreme 
$^3$He enhancement and the unusual shapes of $^3$He spectra. Moreover,
as shown in Petrosian et al.~2009, such a model can  reproduce the large range 
of the observed $^3$He to $^4$He fluence ratio. However,  the pure stochastic
acceleration model cannot explain the  harder
(characteristically broken power-law) spectra observed in the 
gradual events where there is  little or no $^3$He enrichments. Here we explore
the possibility that these harder spectra arise because of re-acceleration in
the CME environment.

Liu et al. (2004) showed that the spectra of He nuclei escaping the coronal
acceleration site consist 
of a low energy quasi-thermal component (which is below the observable 
energy range of $\sim 0.1$ to 10 MeV/nucleon) and a higher
energy nonthermal component with high  $^3$He enrichment. 
Liu et al. (2006)  showed that the
relative strength of the two components depends on many parameters but is most
sensitive to the level of turbulence or  $\tau_p^{-1}\propto f_{\rm turb}$;
 at low levels of
turbulence  a small fraction of $^4$He but most of
$^3$He is accelerated into a nonthermal tail. This  explains the weak
impulsive events with high enrichments. For higher levels of turbulence,
expected
to be the case for larger (and possibly more gradual) events, more of the
quasi-thermal component of $^4$He is moved to higher energies 
so that the ratio of  $^3$He to $^4$He fluences (in the $\sim 0.1$ to 10
MeV/nucleon range) decreases eventually reaching  normal
coronal values. 
However, in higher fluence events  {\it the model spectra} of $^4$He, 
and to some extent that of $^3$He, do
not agree
with the observed spectra of gradual events (As an example see the dashed model
spectra and observed points in Fig. \ref{fig:He} below). Thus, the spectra
coming out of the
coronal acceleration site must be modified by a secondary process. Since high
fluence
events
are more likely to be associated with  fast CMEs, and hence strong shocks,
then this modification can be due to re-acceleration in the  CME environment.
The most attractive feature of this scenario is that even though the nonthermal
tails may be highly enriched, the total number of particles (quasi-thermal plus
nonthermal)
injected into this re-acceleration site can have essentially normal
abundances. It should also be noted that, the shape of these seeds ions
are 
qualitatively similar to the two  (a steep low energy and flatter high
energy) component model used by Tylka \& Lee (2006) in
their phenomenological description  of the  diffusive shock acceleration of
SEPs. However, as mentioned above the origins of these seeds in our scenario is
quite different
than theirs, which assumes previously accelerated particle in the upstream of
the shock. In our scenario the seed particles are  the flare
accelerated ions trapped behind the CME shock. 

\begin{figure}[!ht]
\begin{center}
\includegraphics[width=8.5cm]{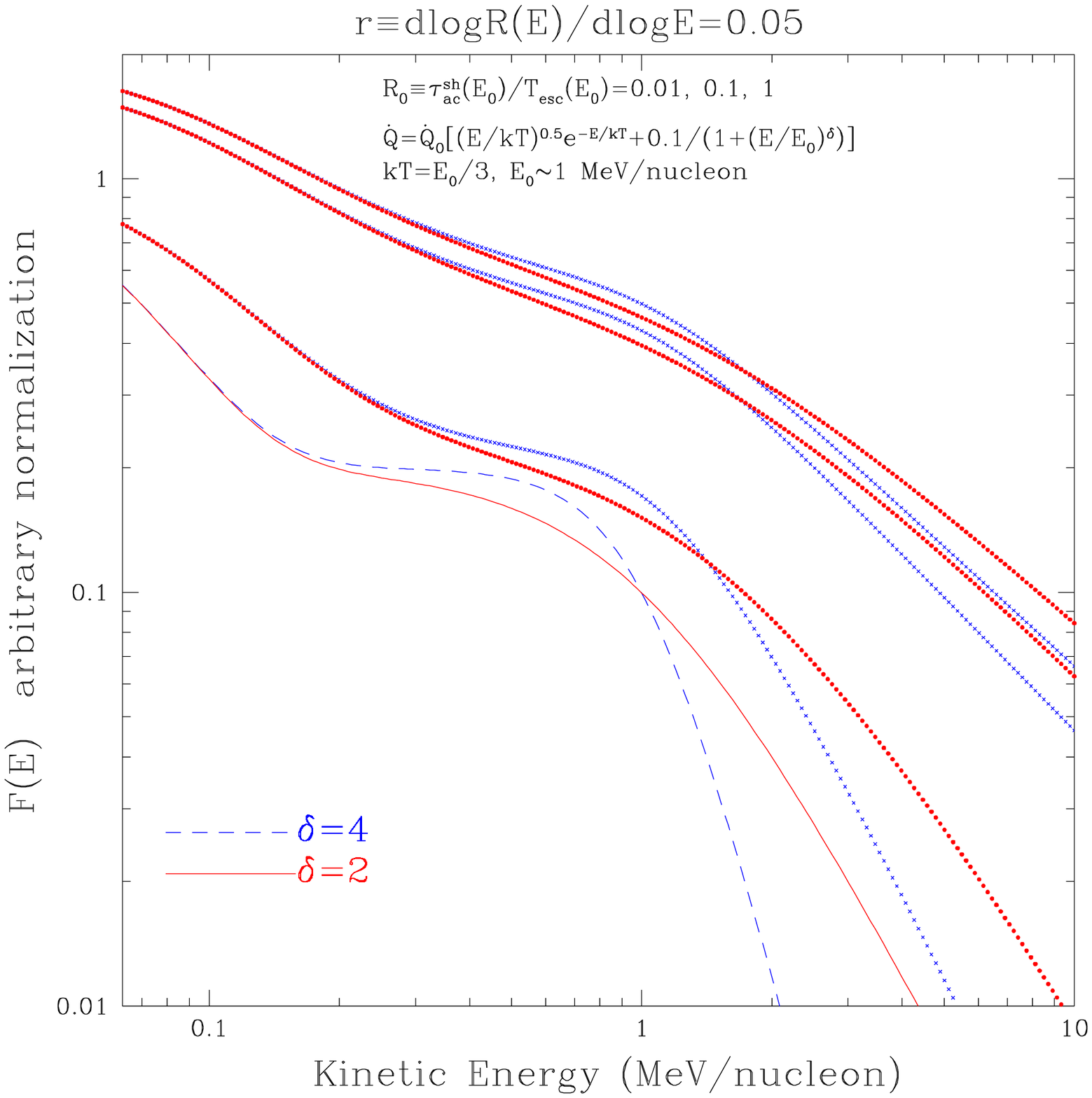}
\includegraphics[width=6.3cm]{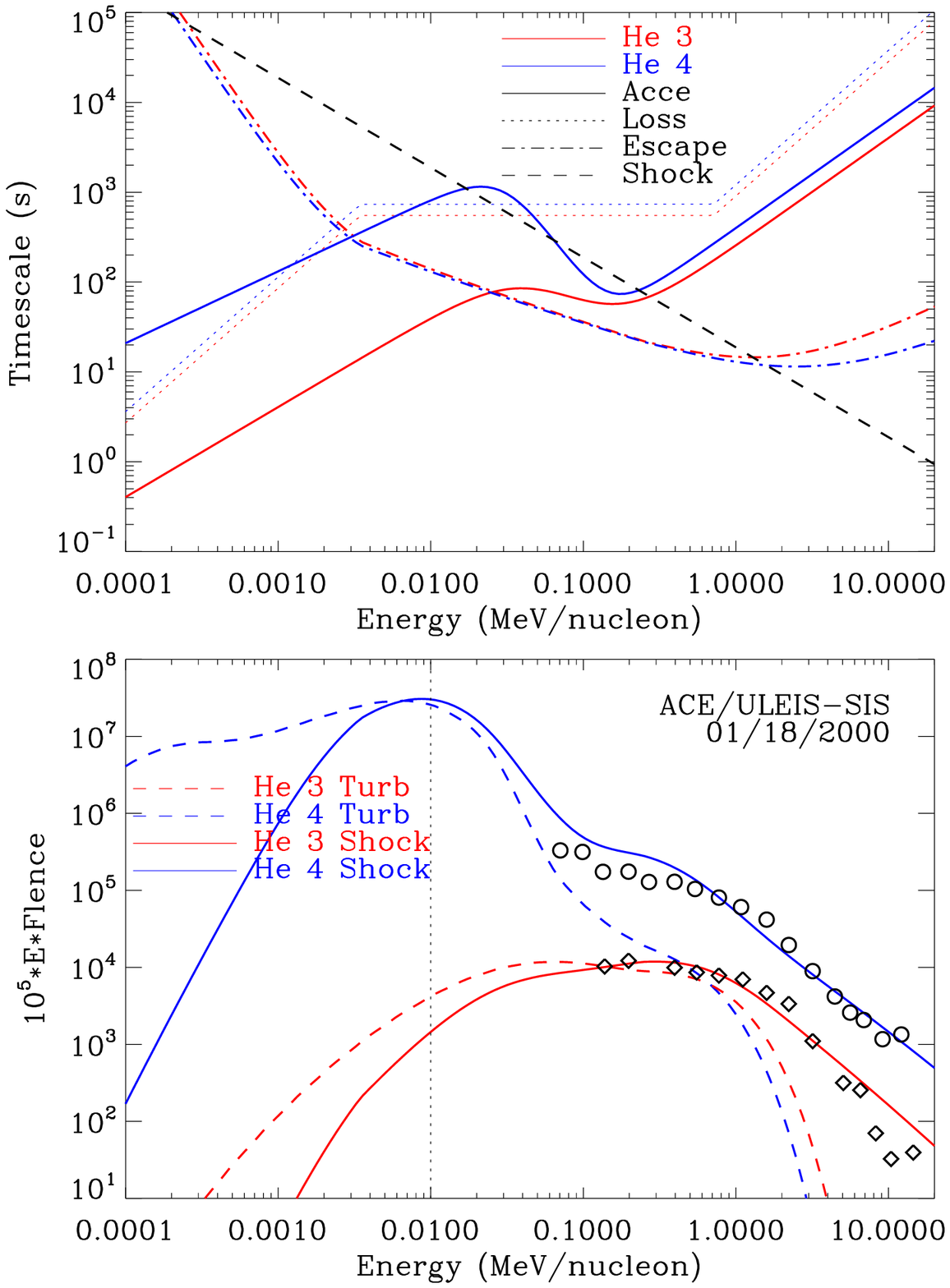}
\caption{{\bf Left:} Spectra  of re-accelerated ions based on
Equation
(\ref{KE'}) with a source term ${\dot Q}'$ consisting of a thermal and a
power-law component with the specified parameters  with two spectral
indexes $\d=2$ (red-solid and dots) and $\d=4$
(blue-dashed and crosses), and
for three values of the ratio ${\cal R}_0$ from top to bottom.  Note 
that for smaller values of ${\cal R}_0$ the spectra are modified into a
broken power law form observed for $^4$He and many other ions. {\bf Right:}
Fluence of $^4$He (blue) and $^3$He (red). The injected spectra
from the flare acceleration site (from Liu et al. 2006) are shown by the dashed
curves 
and the spectra of the re-accelerated ions are shown by the solid curves. The
model
timescales of the the stochastic plus shock acceleration in the CME environment
are shown on the top panel. Stochastic acceleration timescales are from Liu et
al. (2006) to  which we have added an ad-hoc acceleration rate that dominates at
energies $>0.5$ MeV /nucleon, presumably due to a shock.} 
\label{fig:He}
\end{center}
\end{figure}

The equations describing the re-acceleration here is similar to those used above
(Eqs. \ref{KE'} and \ref{solution}) for
electrons. Again the losses can be ignored and we must be dealing with
a strong diffusion case, but  now the injected spectrum ${\dot Q}^\prime$ is
not a simple power law but consists of a quasi-thermal plus a  nonthermal 
component. Figure \ref{fig:He} (left) shows  examples of re-accelerated
spectra (red and blue points) from a simple shock (and/or stochastic
acceleration) model with an injected thermal-plus-power-law  spectrum
(blue-dashed and solid-red curves) similar to those expected to emerge from the
coronal acceleration site (see Liu et al. 2006). As evident, for reasonable
values of the temperature of the quasi-thermal component and power-law indexes
simple broken power laws emerge from the re-acceleration site. This demonstrates
that a generic re-acceleration model can produce spectra similar to those
observed in gradual events. However, a more accurate calculation using realistic
parameters based on the conditions in the environment of the CME is required.
Unfortunately these condition are poorly known. Nevertheless, as a more specific
example in Figure \ref{fig:He} (right) we show a comparison between observation
of a gradual SEP event and result from numerical solution of the full kinetic
Equation (\ref{KE}) of re-accelerated $^4$He and $^3$He spectra using an
examples of flare site accelerated spectra (from Liu et al. 2006) for the
source term ${\dot Q}^{\prime}(E)$. Here in addition to the diffusion term due
to
interaction with turbulence we have added a simple direct acceleration
term, presumably by a shock that dominates at energies $>0.5$ MeV
/nucleon. As evident we get reasonable agreement. These are promising
result but clearly more detail analysis is required to support this scenario.

\section{SUMMARY AND DISCUSSION}
\label{sec:disc}

Several observations comparing the temporal and spectral characteristics of
electrons responsible for the generation of solar flare nonthermal radiations
(type III radio
and HXRs) with those of electrons seen as SEPs near the Earth indicate that the
population can be roughly divided into two groups. In one group, referred to as
``prompt" events, the SEPs appear to originate almost simultaneously with the
radiations from the flare site  located in low corona near the tops of
reconnecting magnetic loops. HXR and SEP observations of a sample of flares
analyzed by
K07 shows that the HXR photon spectral index $\g_{\rm HXR}$ and the SEP
spectral index $\d_{\rm SEP}$ are strongly correlated with the average relation 
$\d_{\rm SEP}\sim \g_{\rm HXR}$. This relation does agree with either
a thin or thick target model of the bremsstrahlung emission. Similarly, the
total (time integrated) number of electrons required for production of HXRs
(based on the thick target model) $N_{\rm HXR}$ appears to also correlate
(though somewhat more weakly) with  the total number of observed SEP electrons
$N_{\rm SEP}$ but the former is larger than the latter  by 100 to 3000 times
while
the simplest model of common origin would imply  comparable numbers.
The second group referred to as ``delayed" events show a complex
temporal relation, often with the deduced time of emission of SEPs at the Sun
coming after that of HXRs or type III radio. K07 data on 15 such  events
shows no or very weak correlation between the two indexes with the SEP index 
being smaller for all except one of the  events. There is no information given
on the relative number of electrons in the delayed events

A similar dichotomy  seems to be present in the observations of SEP ions.
Shorter, weaker events, often referred to as ``impulsive" appear to have higher
enrichment of $^3$He and heavier than CNO ions and softer spectra with an
unusual convex spectral shape for enriched ions. While longer duration and
stronger events show near normal abundances and a harder broken power law
spectra.
Both the delayed  electron and gradual-normal abundance ion events are more
likely to be associated with a fast CME.

In this paper we address  these dichotomies using the scenario that in the
prompt and impulsive events particle acceleration takes predominately at the
reconnecting coronal sites. But in the delayed-gradual events not only there is
acceleration in the flare site but that these accelerated particles are
re-accelerated in the CME-shock environment. This is different than the
assumption made in most models of acceleration by a CME shock, where the seed
particles are some yet unobserved nonthermal component of the upstream
plasma. We quantify this model and show
how such observations can be used to set constraints on the physical
charcteristics of the acceleration mechanism(s).

Our results can be summarized as follows:

\begin{itemize}

\item

We first emphasize that the relation between radiation producing and
SEP electrons is not simple so that the fluxes, numbers  and spectra can not be
assumed to be the same. In particular high resolution spectral and spatial HXR
observation
indicate that flare site acceleration is most likely located in the
reconnection region just above the newly reconnected closed flare loops and the
spectra and the rate of the  escape of these particle to lower atmosphere where
they produce the HXRs could be different than that of the electrons escaping
along open field lines to the Earth. Thus, a more careful analysis  taking such
differences into account is needed.

\item

We show that we can quantitatively account for the above mentioned  observations
using the
simple leaky box
model of acceleration, treating  acceleration by plasma turbulence and
shock  at coronal reconnection and CME sites. The transport
coefficients of this equation are related to two of the most basic
characteristics common in all acceleration model, namely the momentum (or
energy) and pitch angle diffusion coefficients and the background plasma
charcteristics (density, temperature, magnetic field strength and geometry).
We convert the transport coefficients to their characteristics timescales;
energy loss, energy gain or acceleration, energy
diffusion times, and an escape time which depends on the scattering time,
crossing time (across the acceleration site) and the magnetic field geometry.
We describe  an approximate but reliable method of quantifying  these
relations (Fig. \ref{fig:tesc}). For the narrow range of electron energies of
interest here (dictated by the limitations of the observations) we approximate 
the energy dependences of these time scales by power laws.

\item

Using the  K07 data we show that the difference between delayed and prompt
event spectral relations arises primarily because the differences in the SEP
spectra and not the HXR spectra (Fig. \ref{fig:hist}); SEP spectra of delayed
events are harder.

\item

Using the  power law approximation we derive a simple relation
between HXR and SEP indexes (Eq. \ref{dSEP}) from which we derive the expected
distribution of the index $\a_d$ of downward escaping time (Fig.
\ref{fig:alphad})
and show that the two flares for which CP13 obtained this index directly and
non-parametrically from inversion of \r data agree very well with this
distribution. This strengthens the conjecture of CP13 that in the flare 
acceleration site we are dealing with a weak spatial  diffusion and a
relatively strong field convergence in the downward direction.

\item

We also show that in this scenario the number  of electrons required for
production of HXRs is larger than those observed near the Earth (mainly) by the
ratio of energy loss time to downward escape time in the acceleration site. As
shown in CP13 this ratio is much larger than one and depends on the one unknown
parameter which is the density of electrons in the acceleration site. The
observed ratio of $100-3000$ indicates a density of about $\sim 10^9$
cm$^{-3}$ which about 10 times smaller that one gets from the plasma emission
measure obtained from fitting to low energy thermal HXRs.

\item

We then consider the re-acceleration in the CME-shock region (due to both
stochastic and shock accelerations) and using analytic and numerical solutions
of the kinetic equation to show that the harder SEP indexes of delayed events is
consistent with this scenario and point how such observations (of individual and
a population of events) can be used to constrain the re-acceleration rate and
the escape time of SEPs from the CME-shock region. Using calculations based on
interaction of electrons with waves of all modes propagating along the magnetic
field lines we show that for the semi-relativistic electrons of interest
here, stochastic acceleration by turbulence is more
efficient than the expected low Alfv\'en Mach number shocks.

\item

Finally, we briefly consider a similar scenario to account for the differences
between spectra and abundances of SEP $^3$He and $^4$He ions. Based on spectra
obtained in the stochastic acceleration model of Liu et al. (2004 and 2006) we
show that in general re-acceleration of such spectra can produce the
characteristics harder broken power-law spectra seen in gradual events. Also
using
detailed numerical solution we fit the results from such a model to the
observed spectra in one gradual event. This fit requires both diffusion and
acceleration by turbulence and acceleration by a shock which becomes dominant at
higher energies (Fig. \ref{fig:He}).

\end{itemize}

In conclusion then we have demonstrated that with the scenario involving
combined acceleration at the flare site and the CME environment we can explain
some of the puzzling variations seen in spectra of SEP electrons and He ions 
and radiating electrons quantitatively. These result indicate that with more
data and further more rigorous theoretical treatment we can begin to constrain
some of the most important characteristics of the
acceleration mechanism(s) in solar flares and answer many other questions
that these result raise. For example, this scenario requires near
simultaneous launch of a CME  and acceleration  at the reconnection site, and/or
the trapping of the accelerated particles so that they interact  with the CME.
However, since number of SEP
electrons are smaller that the radiating ones we require the trapping behind the
CME shock of only a small fraction of flare accelerated electrons.  Clearly
the geometry of the magnetic fields extending from the reconnection site to the
CME  and characteristics of the turbulence generated behind the CME shock
play  important roles in this process. In general, all shock acceleration
models require presence of turbulence both upstream and downstream of the
shocks and some numerical simulations (see e.g. Giacalone \& Jokipii 2007) show
some indication of the presence turbulence in the downstream region.
These are important aspects that require extensive numerical simulations which
are beyond the scope of this paper.

Acknowledgements: I would like to thank former graduate students, John Leach,
James McTiernan and Qingrong Chen and Post doctoral fellow Siming Liu whose
work contributed to the foundation of the results presented here. This work is
supported by NASA  LWS grant NNX13AF79G, H-SR grant NNX14AG03G and Fermi-GI
grant NNX12AO78G. I would also like to thank an anonymous referee for some
insightful suggestions, G. Zank for pointing out the earlier work by
Forman et al., and F. Effenberger for valuable comments.

\appendix

\section{ACCELERATED ELECTRON SPECTRA FROM TOTAL HXR SPECTRA}

For spatially unresolved  flares we have only the total HXR spectrum $J_{\rm
tot}(\e)$ from
which we can obtain the total spectrum of radiating electrons $N_{\rm
tot}(E)$ either by a forward fitting or by an inversion method (see Kontar et
al. 2005)
of an
equation similar to Equation (\ref{LTspec}). As described in \S \ref{sec:brem}
the relation between this spectrum and that of the accelerated electrons $N(E)$
is not straightforward, especially at lower energies when there could be a
significant emission from both loop top and footpoint sources.  Here we
demonstrate that, in
principal, we can obtain the accelerated spectrum from $N_{\rm tot}(E)$ if we
have a knowledge of the energy dependence of
the escape and energy loss times.

Normally $J(\epsilon)$ is considered to be a thick target spectrum from a single
source. However,  a more accurate
description is that
$J_{\rm tot}(\epsilon)=J_{LT}(\epsilon)+J_{FPs}(\epsilon)$ is the sum of the
{\it thin
target} LT and the {\it thick target} emission from the loop outside
the LT region, mainly from the FPs, with (volume integrated) electron
spectra $N(E)$ and $N_{\rm eff}(E)$ (Eq. \ref{Neff}), respectively. In other
words the deduced total electron spectrum 
\beq\label{totalspec}
N_{\rm tot}(E)=N(E)+{1\over \dot{E}_{\rm L}}\int_E^\infty
{N(E^\prime)\over T_{\rm esc}(E^\prime)}d E^\prime.
\eeq
The fact that $v\dot{E}_L$ is  constant the differentiation of
the above integral equation yields the differential equation 
\beq\label{diffeq}
{d (N/v)\over d E}-{1\over \dot{E}_L T_{\rm esc}}{N\over v} =
{d (N_{\rm tot}/v)\over d E},
\eeq
the solution of which gives the accelerated spectrum%
\footnote{Just a reminder that because the thick target
footpoint emission is independent of density the energy loss time is calculated
here using the density
$n_{\rm lT}$ of the loop top acceleration site.} 
\beq\label{integ}
N(E) = N_{\rm tot}(E) -v\int_{\eta(E)}^\infty{N_{\rm tot}\over
v^{\prime }}e^{\eta-\eta^\prime} d \eta^\prime,
\,\,\,\, {\rm with} \ \ \ \ \
d \eta = {d E \over\dot{E}_{L}T_{\rm esc}}={\tau_L\over T_{\rm esc}}{dE\over
E},
\eeq
with the second term  giving the effective spectrum of 
Equation
(\ref{FPspec}). Clearly the primary charcteristics affecting this  difference is
the relation between $\eta$ and $E$ or the ratio ${\cal R}'=\tloss(E)/\tesc(E)$,
which
is  proportional to $vE/\tsc$ or $v^3E\tsc$ in the weak and strong diffusion
limits, respectively. 

There are, however, some caveats. The simplicity of the above equations is
somewhat deceiving. In equation (\ref{diffeq}) we are required to take a
derivative of the spectrum obtained from inversion of noisy data, which can
limit the accuracy of the  results.%
\footnote{Results obtained by inversion methods
show that the errors in  $N_{\rm tot}$ increase near sharp features (see Brown
et al.~2006).}
Note that
even though Equation (\ref{integ}) does not involve differentiation but
integration  (that tends to smooth out noise),  it involves the subtraction of
two nearly equal (noisy) terms which can amplify the noise. Nevertheless, this
method is promising and with some judicious smoothing reasonable results can be
obtained. Figure \ref{fig:inversion} shows an example where we obtain the
accelerated spectrum $N(E)$ from the total spectrum  obtained by Kontar et
al.~(2005) using  regularized inversion of a \r\, HXR spectrum, using the above
equations and several energy (or velocity) dependence of $\tesc$. The
difference $N_{\rm tot}-N(E)$ gives the effective spectrum $N_{\rm eff}(E)$.
These then can be used to determine the other transport coefficients such as
$D_{EE}$ etc using the formalism described in CP13 (e.g. their Eq. (18)).
\begin{figure}[!ht]
\begin{center}
\includegraphics[width=0.8\textwidth]{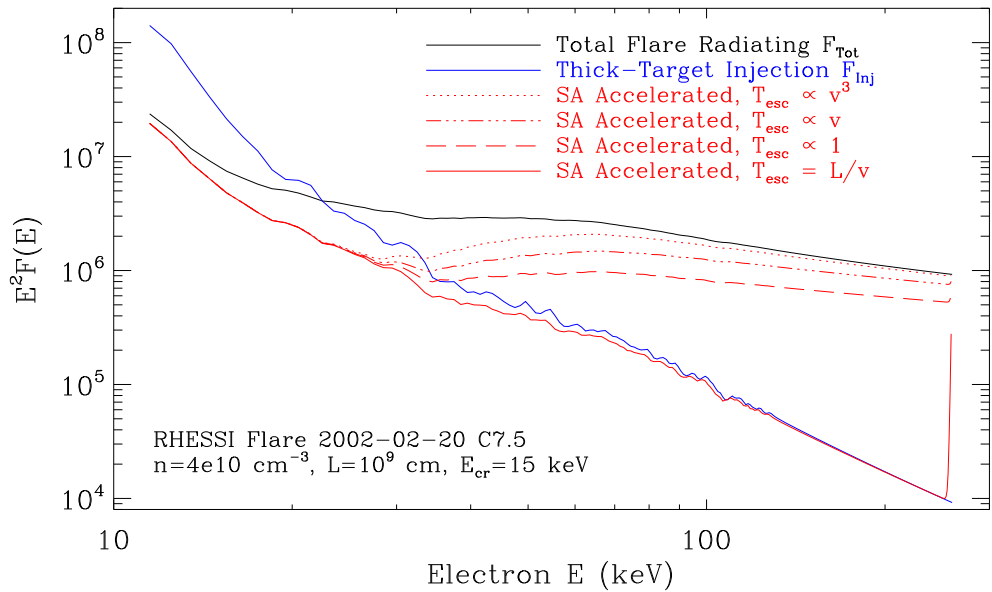}
\caption
{Demonstration of converting the total spectrum of the  radiating
$\overline{N}(E)=N_{\rm tot}(E)$ (solid black curve, from Kontar et al.~(2005))
to the accelerated spectrum $N(E)$ (shown by red solid, dotted and dashed
curves) using Equation (\ref{integ}) and various forms for the escape time.
For comparison we also show the form of the injected spectrum assuming a
simple thick-target model (steep blue solid curve). The curves for $N(E)$ can
be used to obtain the energy dependences of the other transport coefficients
such as $D_{EE}$ and $A(E)$ as demonstrated in CP13 (see also Chen 2013 PhD
Thesis Stanford University).
}
\label{fig:inversion}
\end{center}
\end{figure}

\section{STOCHASTIC RE-ACCELERATION BY TURBULENCE}

For stochastic acceleration by turbulence  the diffusion and direct
acceleration rates are comparable. Then ignoring the loss term the kinetic
Equation \ref{KE} can de written as 
\beq\label{SA}
{d\over dE}
\left[D_{\rm EE}\left({d N\over d E}-{N\over E}\xi
\right)\right]-{N\over \tesc}=\dot{Q}.  
\eeq
If we define an effective acceleration rate 
\beq\label{Aeff}
A_{\rm eff}= (\xi-d\log N/d\log E)D_{EE}/E
\eeq
and assume that the slowly varying logarithmic derivative is constant then
we get similar re-acceleration equation as for shock acceleration with this
effective SA rate replacing $A^\prime$ in Equation (\ref{KE'}) and same
solutions as those presented in Figure \ref{fig:sol} with the acceleration time 
\beq\label{SAacc}
\tau_{\rm ac}^{\rm SA}=E/A_{\rm eff}=\zeta^\prime\tac \,\,\,\,\,
{\rm where}\,\,\,\, \zeta^\prime=\xi^\prime/(\xi-d\log N/d\log E),
\eeq
and $\tac=p^2/D_{pp}$ is the acceleration times calculated in PP97 and used
in the text. As shown in Petrosian \& Chen (2014)
$\xi^\prime=(\g+1)(2\g^2-1)/\g^3\sim 2$
for all energies. For sub-relativistic regime under consideration here
$\xi\sim 0.5$ and from observations we estimate $d\log N/d\log E\sim -1 \,
{\rm to }\, -3$
so
that the correction coefficient  $\zeta^\prime$  varies  between 
1.33 to 0.8.

\section{RE-ACCELERATED SPECTRA FOR $r=0$}

Many treatment of the  acceleration use the simplifying assumption  of energy
independent escape and acceleration times or rate) which is an special case of
when the acceleration and escape time have similar energy dependence and the
ratio ${\cal R}_0$ is constant. This gives $\eta={\cal R}_0\ln (E/E_0)$ and 
 $e^\eta=(E/E_0)^{\cal R}_0$. Substituting this in Equation
(\ref{KE'}) it is easy to show that for a power law injected spectrum
${\dot Q}(E)={\dot Q}_0(e/E_0)^\d$ we get 
\beq\label{solr0}
F(E)={{\dot Q}_0{\cal R}_0\over \d-{\cal R}_0-1}(x^{-{\cal
R}_0-1}-x^{-\d})+F(E_0),\,\,\,\,\,{\rm where}\,\,\,\,\, x=E/E_0.
\eeq
For ${\cal R}_0> \d-1$, i.e. when the acceleration time is longer than
the escape time the spectrum asymptotically approaches the injected
spectrum $F\propto x^{-\d}$), and in the opposite case of ${\cal R}_0<
\d-1$, i.e. shorter acceleration time, $F\propto x^{-{\cal R}_0-1}$ so that
the re-accelerated spectrum becomes harder (index changing by $\d-{\cal
R}_0-1$). Thus, for typical index change of 1 we need  ${\cal R}_0=1 (2)$ for
injected index $\d=3 (4)$.

\bibliographystyle{apj}
\bibstyle{aa}

%\input{XXX.bbl}
%\bibliography{~/Research/Thesis-PhD/Front-Back/ref-bibtex}

%\newpage
%\twocolumn
\def\refer { \par \noindent \hangindent=2pc \hangafter=1}
\baselineskip = 10 true pt
%\section*{REFERENCES}
%\begin{multicols}{2}

\section*{REFERENCES}

\refer Ackermann  Ajello, M., Allafort, A., et al.\ 2012, \apj, 745, 144
\refer Ackermann, M., Ajello, M., Albert, A, et al.\ 2014 \apj, 787, 15
\refer Ajello, M., Albert, A., Allafort, A. et al.  2014 \apj, 789, 20
\refer Brown, J. C. 1972, \solphys, 25, 118
\refer Brown, J. C. et al. 2006, \apj,  643, 523
\refer Chen, Q. 2013, PhD Thesis, Stanford University 
\url{https://purl.stanford.edu/mh246zj4146}
\refer Chen, Q., \& Petrosian, V.\ 2013, \apj, 777, 33;  ({\bf CP13})
\refer Chupp, E. L., Forrest, D. J., Ryan, J. M.  et al. 1982, \apj, 263, L95
\refer Chupp, E. L., \& Ryan, J. M. 2009, Res. in Ast \& Astrophys. 9, 11
\refer  Drake, J.~F., Swisdak, M., Che, H., \& Shay, M.~A.\  2006, \nat, 443,
553
\refer Drake, J.~F., Swisdak, M.,  \& Ferno, R. \ 2013, \apjl, 763, L5
\refer Fisk, L.~A.\ 1978, \apj, 224, 1048
\refer Forman, M. A., Webb, G. M., \&  Axford, W. I. 1981, ICRC, 9, 238
\refer Giacalone, J., \& Jokipii, J. R. \ 2007, \apjl, 663, L41
\refer Gopalswamy, N., \&  Yashiro, S.\ 2011, \apj, 736, L17
\refer Guo, F. \& Giacalone, J., \ 2012, \apj, 753, 28
\refer Haggerty, D. K., \&  Roelof, E. C. 2002, \apj, 579, 841
\refer Ho, G. C., Roelof, E. C., \& Mason G. M. 2005, \apjl,  621, L141
\refer Hurford, G. J.,  Schwartz, R. A.,  Krucker, S., Lin, R. P., Smith, D.
M. \& Vilmer, N. 2003, \apjl, 595, L77
\refer Jokipii, J. R., \& Giacalone, J.\ 1996 Space Sci. Rev., 78, 137  
\refer Klein, K.-L., Krucker, S., Trottet, G., \&  Hoang, S. 2005, A\&A, 431, 10
\refer Koch, H. W. \& Motz, J. W. 1959, Rev. Modern Phys., 31, 920  
\refer Kontar, E. P., Emslie, A. G., Piana, M. et al. 2005, \solphys,
226, 317
 M.\refer Krucker, S., Larson, D. E.,  Lin, R. P., \& Thompson, B. J. 1999,
\apj,
519, 864
\refer Krucker, S., Kontar, E. P., Christe, S., \& Lin, R. P. 2007, \apjl, 663,
L109 ({\bf K07})
\refer Krucker, S., \& Lin, R.~P.\ 2008, \apj, 673, 1181
\refer Le Roux, J. A., Zank, G. P., Webb, G. M., \& Khabarova,  2015,
\apj,  801, 112 
\refer Le, A., Karimabadi, H., Egedal, J., Roytershteyu, v., \& Daughton, W. \
2012, Plasma Phys., 19, 072120
\refer  Lee, M.A., \apjs, 158, 38, 2005
\refer Lin, R. P., Dennis, B. R., \& Hurford, G. J. 2002, Sol. Phys., 210, 3
\refer Lin, R. P., \& Hudson, H. C. 1971, \solphys, 17, 412
\refer Liu, S., Petrosian, V., \& Mason, G.~M.\ 2004, \apjl, 613, L81
\refer Liu, S., Petrosian, V., \& Mason, G.~M.\ 2006, \apj, 636, 462
\refer Liu, W., Petrosian, V., Dennis, B.~R., \& Jiang, Y.~W.\ 2008, \apj, 676,
704 
\refer {Liu}, W., {Chen}, Q., \& {Petrosian}, V. 2013, \apj, 767, 168
\refer Maia, D. J. F., \&  Pick, M. 2004, \apj, 609, 1082
\refer Mason, G.~M., Reames, D.~V., von Rosenvinge, T.~T., Klecker, B., 
  \& Hovestadt, D.\ 1986, \apj, 303, 849 \refer Malyshkin, L.  \& Kulsrud, R.
2001, \apj, 549, 404
\refer Mason, G. M., Dwyer, J. R., \& Mazur, J. E. 2000, \apj 545, 157L
\refer Mason, G. M. et al. 2002, \apj 574, 1039
\refer  Masuda, S., Kosugi, T., Hara, H., Tsuneta, S., \& Ogawara, Y.\ 1994,
\nat, 371, 495
\refer Mazur, J. E. et al. 1992, \apj 401, 398
\refer Melissa Pesce-Rollins, Nicola Omodei, Vah\'e  Petrosian, Wei Liu,
Fatima
Rubio da Costa, Alice Allafort, Qingrong Chen, 2015 \apjl, 805. 15
\refer Miller, J. A., LaRosa, T.N., \& Moore, R.L. 1996, \apj 445, 464
\refer Miller, J. A.\ 2003, {\it COSPAR Colloquia Series} Vol. 13, 387
%\refer Miller, J.~A., \& Reames, D.~V.\ 1996, American Institute of Physics
%Conference Series, 374, 450
\refer Ng, C.~K., \& Reames, D.~V.\ 1994, \apj, 424, 1032 
\refer Nitta, N.~V., Freeland, S.~L., \& Liu, W.\ 2010, \apjl, 725, L28 
\refer Oka, M., Phan, T.-D., Krucker, S., Fujimoto, M. \& Shinokava, I.\ 2010,
\apj, 714, 915
\refer Parker, L. N., \&  Zank,  G. P.\ 2012, \apj, 757, 97
\refer Petrosian, V. 1973, \apj, 186, 291
\refer Petrosian, V. 1982, \apjl, 255, L85
\refer Petrosian, V. 2012, Space Sci Rev., 173, 535
\refer Petrosian, V., \& Kang, B. 2015, \apj, 813, 5
\refer {Petrosian}, V., \& {Donaghy}, T.~Q. 1999, \apj, 527, 945
\refer Petrosian, V., Donaghy, T.~Q., \& McTiernan, J.~M.\ 2002, \apj, 569, 459
\refer Petrosian, V., \& Liu, S.\ 2004, \apj, 610, 550
\refer Petrosian, V., Jiang, Y.~W., Liu, S., Ho, G.~C., \& Mason, G.~M.\ 2009,
\apj, 701, 1
\refer Piana, M., Massone, A.~M., Kontar, E.~P., et~al. 2003,
\apjl, 595, L127
\refer Pryadko, J.~M., \& Petrosian, V.\ 1997, \apj, 482, 774 ({\bf PP97}) 
\refer Pryadko, J.~M., \& Petrosian, V.\ 1998, \apj, 495, 377
\refer Pryadko, J.~M., \& Petrosian, V.\ 1999, \apj, 515, 873
\refer Ramaty, R., \& Murphy, R.~J.\ 1987, \ssr, 45, 213 
\refer Reames, D. V., Meyer, J. P., \& von Rosenvinge, T. T.\ 1994, ApJS, 90,
649
\refer Reames, D. V. 2013, Sp. Sci. Review, 175, 53refer Reames, D. V. et al.
1997, \apj 483, 515
\refer Reames, D. V., Cliver, E. W., \& Kahler, S. W. 2014, \solphys, 289, 4675
\refer Russell, R. T., \& Mulligan, T. 2002, Planet. Space Sci., 50, 52
\refer Schlickeiser, R.\ 1989, \apj, 336, 243
\refer Steinacker, J., Schlickeiser, R., \& Droge, W.\ 1988, \solphys, 115, 313 
\refer Tylka, A.~J., \& Lee, M.~A.\ 2006, \apj, 646, 1319 
\refer Wang, L., Lin, R.~P.,  Krucker, S., \& Mason, G.\ 2012, \apj, 759, 69
\refer Zank, G. P., Le Roux, J. A., Webb, G. M., Dosch, A., \&Khoborova, d. \
2014, \apj, 797, 28
\refer Zank, G. P., Hunana, P.,  Mostafavi, P., Le Roux, J. A., Li, Gang,
Webb, G. M., Khabarova, O., Cummings, A., Stone, E., \& Decker, R, 2015 \apj,
814, 28

%\end{multicols}

\end{document}